\documentclass[journal]{IEEEtran}
\usepackage{amssymb}

\usepackage{cite}

\usepackage{hyperref}

\usepackage{dirtytalk}

\ifCLASSINFOpdf
\else
\fi
\usepackage{stackengine}
\usepackage{amsmath}
\usepackage{amsfonts}
\usepackage{siunitx}

\usepackage{array}
\makeatletter
\let\MYcaption\@makecaption
\makeatother

\usepackage{silence}
\usepackage[font=footnotesize]{subcaption}

\makeatletter
\let\@makecaption\MYcaption
\makeatother

\usepackage{multirow}

\usepackage{amsthm}
\usepackage{graphicx}
\graphicspath{ {./images/} }

\usepackage[section]{placeins}
\usepackage{float}

\usepackage{pgfplots}
\pgfplotsset{compat=newest}
\usetikzlibrary{calc}
\usetikzlibrary{plotmarks}
\usetikzlibrary{arrows.meta}
\usetikzlibrary{external}
\usetikzlibrary{decorations.shapes} 
\usepgfplotslibrary{patchplots}
\usepackage{grffile}
\usepackage{amsmath}
\makeatletter
\long\def\ifnodedefined#1#2#3{%
    \@ifundefined{pgf@sh@ns@#1}{#3}{#2}%
}
\makeatother

\usepackage{tikzscale}

\usetikzlibrary{positioning}

\usepackage{imw-tikz}

\usepackage{xcolor}

\definecolor{myorange}{rgb}{0, 0.5, 0}
\definecolor{myturkis}{rgb}{0.8, 0, 1}
\definecolor{myblue}{rgb}{0, 0, 1}

\begin{document}

\DeclareRobustCommand{\lineleg}[1]{\tikz[baseline=1.7ex]\draw[#1, line width=0.4mm] (0,0.3) -- (0.5,0.3);} 

\newcommand{\annotateChangesStart}[0]{}
\newcommand{\annotateChangesEnd}[0]{}

\renewcommand{\annotateChangesStart}[0]{\color{black}}
\renewcommand{\annotateChangesEnd}[0]{\color{black}}







\twocolumn[
\begin{@twocolumnfalse}

\setcounter{page}{0}

\textbf{Author's post-print} \\

\copyright 2026 IEEE. Personal use of this material is permitted. Permission from IEEE must be obtained for all other users, in any current or future media, including reprinting/republishing this material for advertising or promotional purposes, creating new collective works, for resale or redistribution to servers or lists, or reuse of any copyrighted components of this work in other works.\\\\
DOI: \url{https://doi.org/10.1109/TAP.2026.3656171} \\ 


\end{@twocolumnfalse}
]

%
\title{On the Relation of Characteristic Modes of Different Conducting Structures}

%
%
%

\author{Leonardo~M\"orlein,~\IEEEmembership{Student Member,~IEEE,}
        and~Dirk~Manteuffel,~\IEEEmembership{Member,~IEEE}
\thanks{The authors are with the Institute of Microwave and Wireless Systems, Leibniz University Hannover, Appelstr. 9A, 30167 Hannover, Germany (e-mail:
moerlein@imw.uni-hannover.de; manteuffel@imw.uni-hannover.de).}
}

%
%

\markboth{}%
{Mörlein \MakeLowercase{\textit{et al.}}: TBD}
%



\maketitle

\begin{abstract}
A formalism is derived to analyze the scattering of a conducting structure based on the characteristic modes of another structure whose surface is a superset of the first structure. This enables the analysis and comparison of different structures using a common basis of characteristic modes. Additionally, it is shown that the scattering matrices and perturbation matrices are no longer diagonal in these cases. Based on this, a modal transformation matrix is defined to describe the mapping between the characteristic fields and the weighting coefficients of the two structures. This matrix enables the conversion of the perturbation matrices in different bases. Finally, \annotateChangesStart three \annotateChangesEnd examples are provided along with a discussion of some aspects of the theory. The first \annotateChangesStart two examples \annotateChangesEnd aim to validate and illustrate the formalism. The \annotateChangesStart third \annotateChangesEnd example shows how the formalism can be applied in the design process of an antenna element that is gradually modified, starting from a base structure.
\end{abstract}


%
\IEEEpeerreviewmaketitle

\section{Introduction}

In recent years, characteristic modes (CMs) \cite{garbacz_generalized_1968,harrington_theory_1971,sarkar_expose_2016} have become a widely used tool in the design of antenna systems \cite{adams_antenna_2022}. 
Since CMs provide an eigensolution to the electromagnetic scattering fields of a particular structure, they are a suitable decomposition to investigate any kind of antenna or scattering problem of this structure. This decomposition therefore often provides intuitive insight into the scattering phenomena, leading to a diverse field of applications ranging from platform-based antennas \cite{ma_design_2019,li_synthesis_2022,grundmann_cupola-shaped_2025}, over MIMO antennas \cite{kim_systematic_2018,peitzmeier_systematic_2019} to antenna arrays \cite{morlein_beamforming_2021,manteuffel_characteristic_2022} and many more.

The goal in antenna design is usually to find a structure that is suitable to achieve the desired antenna parameters. This means that the structure is, of course, subject to change during the design process. It is a common practice to e.g. cut slots into the \annotateChangesStart initial \annotateChangesEnd antenna shape, to add metal, to add ports or lumped impedances, etc. As the CMs are defined to diagonalize the scattering operator of the structure, this means that for every modification the antenna designer does, the characteristic currents, fields and eigenvalues change. However, this is usually not a problem when CMs are used to obtain intuitive insight into the structure and the modifications are made by an antenna designer based upon this intuition. E.g. in \cite{lin_design_2022}, slots are cut into a patch antenna and the CMs of the old and new structure are treated as if they were equal but just with different eigenvalues. \annotateChangesStart This works well within their methodology, since the modal current distributions only differ slightly when gradual changes are applied and the CMs are only used for intuition. Therefore no strict association between the modes of the structures is necessary. \annotateChangesEnd



However, when CMs are used as a mathematical tool, where not only intuition is necessary but also a computational relation is needed, the former approach is not always sufficient. Recently, we published a paper \cite{morlein_array_2025} where CMs are used as a decomposition for elements in an antenna array. Thereby, the modal scattering parameters in terms of CMs were left as degrees of freedom for an optimizer, representing the geometrical changes during the design process as abstract parameters. Here, it was found that in some cases, the assumption that the characteristic fields do not change significantly w.r.t. geometrical changes can provide sufficiently accurate results. However, as expected, there are other cases \cite{morlein_deembedding_2025}, where the CMs change so much that a new coupling model has to be computed to obtain accurate results. This led to the question if there is a way to account for the changes in the CMs, without recalculating the whole coupling model.

\annotateChangesStart

However, this question is not limited to the array case. It applies to all cases where CMs are used to compare different or modified structures, also for single, isolated elements. Therefore, in this paper, the question is addressed, how the CMs of different or modified structures in general are related to each other, leaving the application to the array case aside.

As basis for this relation, a scattering and perturbation matrix formulation of CMs is well-suited, as it was already introduced in \cite{harrington_theory_1971}. However, in \cite{harrington_theory_1971}, CMs are only used to analyze scattering of the structure whose scattering operator they diagonalize and no relation between the CMs of different structures is given. Also in other publications about characteristic modes since then, no such relation is found.

Similarly, no formalism that could be transferred to CMs is known in other modal decompositions that rely on scattering and perturbation matrix definitions. For example, if spherical-wave functions are considered \cite{hansen_spherical_1988}, they are defined analytically and do not depend on the scattering object. This means that there is also no notion of a transformation between sets of different bases dependent on the object, that could be transferred to CMs.

Therefore, a novel formalism is proposed in this paper that enables to describe scattering problems in the CMs of other structures. \annotateChangesEnd
It can be used not only to treat and compare the scattering of different structures computationally within the same CM basis, but  \annotateChangesStart its introduction also provides \annotateChangesEnd a broader intuition of what CMs are and how they relate to each other.






In section~\ref{sec:definitions}, basic definitions are given. In the sections~\ref{sec:scattering_in_different_decompositions} and \ref{sec:transformation}, the theoretical derivation of the new formalism is given. In section~\ref{sec:examples}, \annotateChangesStart three \annotateChangesEnd examples are given that illustrate the proposed formalism. In section~\ref{sec:discussion}, some aspects of the new formalism are discussed, followed by a conclusion in section~\ref{sec:conclusion}.

\section{Definitions}
\label{sec:definitions}



\subsection{Method of Moments}

Since the focus in this paper is to analyze non-closed perfectly conducting objects, the electric field integral equation (EFIE) formulation of the method of moments is used.

\begin{figure}
    \centering
    \includegraphics{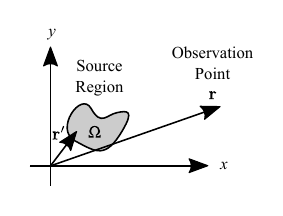}
    \caption{Illustration of the surface of a scattering object $\Omega$, the source position vector $\mathbf{r}^\prime$ and the observation position vector $\mathbf{r}$.}
    \label{fig:coordinates}
\end{figure}

Assuming a scattering object with surface $\Omega$, as seen in Fig.~\ref{fig:coordinates}, discretized by the basis functions $\mathbf{\psi}_m(\mathbf{r}^\prime)$ and the test functions $\widetilde{\mathbf{\psi}}_\mu(\mathbf{r})$, the scattering problem is formulated as:
\begin{equation}
\label{eq:ZI_eq_V}
    \mathbf{Z} \mathbf{I} = \mathbf{V}.
\end{equation}

The impedance matrix $\mathbf{Z}$ is given by: 
\begin{equation}
    \mathbf{Z} = \left[ \int_\Omega\int_\Omega \widetilde{\mathbf{\psi}}_{\mu}(\mathbf{r}) \, \mathbf{G}(\mathbf{r}, \mathbf{r}^\prime) \, \mathbf{\psi}_m(\mathbf{r}^\prime) \mathrm{d}S \mathrm{d}S^\prime \right]_{\mu m},
\end{equation}
whereby $\mathbf{G}(\mathbf{r}, \mathbf{r}^\prime)$ is the dyadic greens function of the background problem.

The right hand side of \eqref{eq:ZI_eq_V} is a tested version of the incident field
\begin{equation}
\label{eq:einc_projection}
    \mathbf{V} = \left[ \int_\Omega \widetilde{\mathbf{\psi}}_{\mu}(\mathbf{r}) \,\mathbf{E}_{\mathrm{inc}}(\mathbf{r}) \mathrm{d}S \right]_\mu
\end{equation}
and the current vector \annotateChangesStart
\begin{equation}
    \mathbf{I} = \begin{bmatrix}
        i_1 \\
        i_2 \\
        ...
    \end{bmatrix}
\end{equation}
contains the coefficients $i_m$ of the basis functions $\mathbf{\psi}_m(\mathbf{r}^\prime)$ in the \annotateChangesEnd current distribution $\mathbf{J}$:
\begin{equation}
    \mathbf{J}(\mathbf{r}^\prime) = \sum_m i_m \mathbf{\psi}_m(\mathbf{r}^\prime).
\end{equation}

\DeclareRobustCommand{\lineleg}[1]{\tikz[baseline=1.7ex,domain=0:0.5] {\node[minimum width=0.5cm, minimum height=0.6em] at (0.25,0.3){};\draw[color=#1,mark repeat=30,mark phase=13,outer sep=1cm,line width=1.5pt] plot (\x,0.3);}}

\subsection{Characteristic Modes}

Characteristic modes provide an eigensolution to the scattering fields of a particular object \cite{garbacz_modal_1965}. There are multiple ways to calculate characteristic modes \cite{harrington_computation_1971,gustafsson_unified_2022-1,gustafsson_unified_2022,capek_characteristic_2023}. Here, the formulation based on an eigenvalue problem of the impedance matrix is used:
\begin{equation}
\label{eq:z0_cm_definition}
    \operatorname{Im}\left\{\mathbf{Z}\right\} \mathbf{I}_{n} = \operatorname{Re}\left\{\mathbf{Z}\right\} \mathbf{I}_{n} \lambda_n,
\end{equation}
whereby $\mathbf{I}_n$ is \annotateChangesStart the representation of the $n$\nobreakdash-th characteristic current distribution $\mathbf{J}_n(\mathbf{r}^\prime)$ in the basis functions $\mathbf{\psi}_m(\mathbf{r}^\prime)$ \annotateChangesEnd of the method of moments as column vector and $\lambda_n$ is the eigenvalue. Throughout this paper, the characteristic modes are normalized such that 
\begin{equation}
\label{eq:cm_normalization}
    \mathbf{I}_{n}^\mathrm{T} \operatorname{Re}\left\{\mathbf{Z}\right\} \mathbf{I}_{n} = 1.
\end{equation}
For the sake of simplicity, a matrix $\mathbf{I}_{\mathrm{CM}}$ is defined that contains all characteristic currents $\mathbf{I}_{n}$ as columns:
\begin{equation}
    \mathbf{I}_{\mathrm{CM}} = \left[\mathbf{I}_{1}, \mathbf{I}_{2}, ... \right].
\end{equation}

Since the formalism in this paper relies on a scattering problem formulation, the characteristic fields $\mathbf{E}_n(\mathbf{r})$ are introduced.
According to \cite{harrington_theory_1971}, the characteristic fields $\mathbf{E}_n(\mathbf{r})$ form a Hilbert space of all fields throughout space produced by currents on $\Omega$. Assuming the real-valued characteristic currents $\mathbf{J}_n$ on $\Omega$ are known, the fields $\mathbf{E}_n(\mathbf{r})$ can be calculated according to:
\begin{equation}
    \mathbf{E}_n(\mathbf{r}) = \int_\Omega \mathbf{G}(\mathbf{r}, \mathbf{r}^\prime) \, \mathbf{J}_n(\mathbf{r}^\prime) \mathrm{d}S^\prime,
\end{equation}
satisfying the wave equation
\begin{equation}
\label{eq:wave_equation}
    \nabla \times \nabla \times \mathbf{E}_n-k^2 \mathbf{E}_n = - \mathrm{j} \omega \mu \mathbf{J}_n.
\end{equation}

While $\mathbf{E}_n(\mathbf{r})$ are of an outward traveling type, standing wave characteristic mode fields $\mathbf{E}^0_n(\mathbf{r})$ can also be obtained according to \cite{harrington_theory_1971} using:
\begin{equation}
\begin{split}
    \mathbf{E}^0_n(\mathbf{r}) &= \frac{1}{2}\mathbf{E}_n(\mathbf{r})\!+\! \frac{1}{2} \mathbf{E}^*_n(\mathbf{r}) \\ &= \int_\Omega \operatorname{Re} \left\{ \mathbf{G}(\mathbf{r}, \mathbf{r}^\prime) \right\} \,\mathbf{J}_n(\mathbf{r}^\prime) \mathrm{d}S^\prime,
\end{split}
\end{equation}
which satisfies the source-free wave equation
\begin{equation}
    \nabla \times \nabla \times \mathbf{E}^0_n-k^2 \mathbf{E}^0_n = 0
\end{equation}
since $\mathbf{E}^*_n(\mathbf{r})$ satisfies the conjugate equation of \eqref{eq:wave_equation}.

With these definitions, the incident field on the surface of $\Omega$ is decomposed into:
\begin{equation}
    \left. \mathbf{E}_\mathrm{inc}(\mathbf{r}) \right|_{\mathrm{tan}} = \sum_n a_n \, \mathbf{E}^0_n(\mathbf{r}) \hspace{3mm}\mathrm{for} \hspace{3mm} \mathbf{r} \in \Omega,
\end{equation}
whereby $a_n$ are the modal excitation coefficients\footnote{\annotateChangesStart In \cite{harrington_theory_1971}, the modal excitation coefficients are called $V_n^i$. Here, we use $a_n$ in order to be consistent with recent literature in the domain of spherical-wave function scattering, as e.g. seen in  \cite{gustafsson_unified_2022-1}, \cite{hansen_spherical_1988}. \annotateChangesEnd} \cite{harrington_theory_1971,grundmann_using_2021}:
\begin{equation}
\label{eq:an_calc_general}
    a_n = \int_\Omega \mathbf{J}_n(\mathbf{r})\, \mathbf{E}_\mathrm{inc}(\mathbf{r}) \,\mathrm{d}S = \mathbf{I}_{n}^\mathrm{T} \mathbf{V},
\end{equation}
and the scattered field $\mathbf{E}_\mathrm{sc}(\mathbf{r})$ is described by the weighting coefficients of the scattered field $f_n$ throughout space according to:
\begin{equation}
    \mathbf{E}_\mathrm{sc}(\mathbf{r}) = \sum_n f_n \, \mathbf{E}_n(\mathbf{r}).
\end{equation}
\annotateChangesStart The incident field coefficient vector $\mathbf{a}$ and the scattered field coefficient vector $\mathbf{f}$ are connected by the perturbation matrix $\mathbf{P}$ according to:
\begin{equation}
\label{eq:fpa}
    \mathbf{f} = \mathbf{P} \mathbf{a},
\end{equation}
whereby $\mathbf{P}$ contains the scattering properties of the scattering object. When the characteristic modes which are used to expand the incident and scattered fields are the eigensolutions of the scattering object in question, the perturbation matrix $\mathbf{P}$ is diagonal and can be calculated from the eigenvalues $\lambda_n$ as stated by Harrington \cite{harrington_theory_1971}:
\begin{equation}
\label{eq:p_diag}
     \mathbf{P} = - \operatorname{diag} \frac{1}{1 + \mathrm{j}\lambda_n}.
\end{equation}

\annotateChangesEnd



\section{Scattering in Characteristic Modes of Another Structure}
\label{sec:scattering_in_different_decompositions}

\begin{figure}
    \centering
    \includegraphics{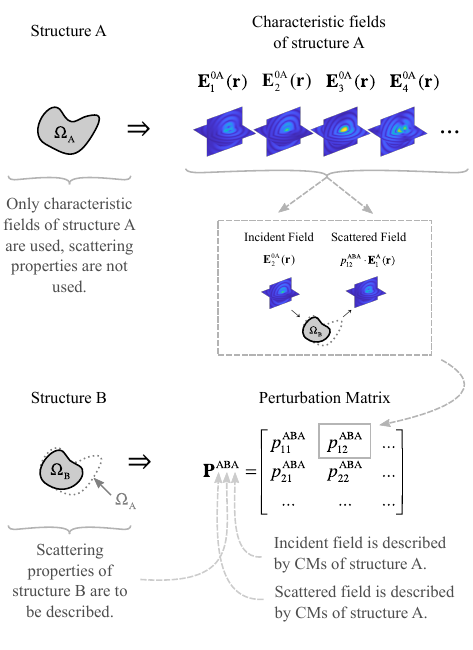}
    \caption{\annotateChangesStart Overview of the construction of the perturbation matrix $\mathbf{P}^{\mathrm{ABA}}$ of a scattering object B in terms of the CMs of another object A.\annotateChangesEnd}
    \label{fig:foreign_structure_CM}
\end{figure}

In order to enable describing multiple structures mathematically in a common CM basis, it is proposed in the following to formulate the scattering problem of one structure in the CMs of another structure. It is assumed that the CMs of structure A are known and the scattering of structure B is to be analyzed in the modes of structure A, as shown in Fig.~\ref{fig:foreign_structure_CM}. \annotateChangesStart As it will be seen throughout the derivation, it is necessary to impose the limitation that the surface of structure~B is a subset of the surface of structure~A. \annotateChangesEnd The superscripts $...^\mathrm{A}$ and $...^\mathrm{B}$ are added to mathematical symbols to indicate if a quantity is originated from structure A, B or both.

The outward traveling characteristic fields of structure A are given by:
\begin{equation}
    \mathbf{E}_n^{\mathrm{A}}(\mathbf{r}) = \int_{\Omega_\mathrm{A}} \mathbf{G}(\mathbf{r}, \mathbf{r}^\prime) \, \mathbf{J}^{\mathrm{A}}_n(\mathbf{r}^\prime) \mathrm{d}S^\prime,
\end{equation}
and standing wave type fields are given by:
\begin{equation}
    \mathbf{E}^{0\mathrm{A}}_n(\mathbf{r}) = \int_{\Omega_\mathrm{A}} \operatorname{Re}\left\{\mathbf{G}(\mathbf{r}, \mathbf{r}^\prime)\right\} \, \mathbf{J}^{\mathrm{A}}_n(\mathbf{r}^\prime) \mathrm{d}S^\prime.
\end{equation}

Using this definition, a formula for the incident field is given as:
\begin{equation}
\label{eq:incident_field_sum_A}
    \left. \mathbf{E}_\mathrm{inc}(\mathbf{r}) \right|_{\mathrm{tan}} = \sum_n a_n^{\mathrm{A}} \, \mathbf{E}^{0\mathrm{A}}_n(\mathbf{r}) \hspace{3mm}\mathrm{for} \hspace{3mm} \mathbf{r} \in \Omega_\mathrm{A},
\end{equation}
whereby $a_n^{\mathrm{A}}$ is the modal excitation coefficient:
\begin{equation}
    a_n^{\mathrm{A}} = \int_{\Omega_\mathrm{A}} \mathbf{J}_n^{\mathrm{A}}(\mathbf{r})\, \mathbf{E}_\mathrm{inc}(\mathbf{r}) \mathrm{d}S.
\end{equation}

\annotateChangesStart Now, to analyze the scattering of structure B in the basis of the modes of A in a method of moments approach, the incident modal fields $\mathbf{E}^{0\mathrm{A}}_n(\mathbf{r})$ are tested with the test functions $\widetilde{\mathbf{\psi}}_{\mu}^\mathrm{B}(\mathbf{r})$ of structure B:
\begin{equation}
    \mathbf{U}^{\mathrm{BA}} = \left[ \int_{\Omega_\mathrm{B}} \widetilde{\mathbf{\psi}}_{\mu}^\mathrm{B}(\mathbf{r}) \mathbf{E}^{0\mathrm{A}}_n(\mathbf{r}) \mathrm{d}S \right]_{\mu n}\!\!\!\!\!\!.
\end{equation}

By introducing the matrix $\mathbf{R}^{\mathrm{BA}}$ that we propose to call cross-radiation matrix as discussed in Appendix~\ref{sec:interpretation_RAB}
\begin{equation}
\label{eq:RBA}
    \mathbf{R}^{\mathrm{BA}}  = \left[\int_{\Omega_\mathrm{B}}\int_{\Omega_\mathrm{A}}\!\!\widetilde{\mathbf{\psi}}_{\mu}^\mathrm{B}(\mathbf{r}) \operatorname{Re}\left\{\mathbf{G}(\mathbf{r}, \mathbf{r}^\prime)\right\} \mathbf{\psi}_{m}^\mathrm{A}(\mathbf{r}^\prime) \mathrm{d}S\mathrm{d}S^\prime \right]_{\mu m}\!\!\!\!\!\!\!\!\!,
\end{equation}
the incident field projection $\mathbf{U}^{\mathrm{BA}}$ can be written as:
\begin{equation}
    \mathbf{U}^{\mathrm{BA}} = \mathbf{R}^{\mathrm{BA}} \mathbf{I}^\mathrm{A}_{\mathrm{CM}}.
\end{equation}

Assuming the modal excitation coefficients with respect to the modes of structure A are known, the right hand side of the method of moments equation $\mathbf{V}^{\mathrm{B}}$ can be written for any $\mathbf{E}_\mathrm{inc}(\mathbf{r})$ as:
\begin{equation}
\label{eq:vb}
    \mathbf{V}^{\mathrm{B}} = \left[ \int_{\Omega_\mathrm{B}} \widetilde{\mathbf{\psi}}_{\mu}^\mathrm{B}(\mathbf{r}) \,\mathbf{E}_{\mathrm{inc}}(\mathbf{r}) \mathrm{d}S \right]_\mu = \mathbf{U}^{\mathrm{BA}} \mathbf{a}^{\mathrm{A}},
\end{equation}
but strictly only if $\Omega_\mathrm{B} \subset \Omega_\mathrm{A}$ since \eqref{eq:incident_field_sum_A} is only valid for $\mathbf{r} \in \Omega_\mathrm{A}$. This means that the surface of structure B has to be a subset of the surface of structure A for this formalism to be exact.

Now, the perturbation matrix of structure B in the basis of the CMs of structure A can be defined as\footnote{\annotateChangesStart A similar formula can be found  for spherical-wave functions in \cite{gustafsson_unified_2022-1}, while the authors call it transition matrix there.\annotateChangesEnd}:
\begin{equation}
\label{eq:p_aba_mom_definition}
\mathbf{P}^{\mathrm{ABA}} = -\left(\mathbf{U}^{\mathrm{BA}}\right)^\mathrm{T} \left(\mathbf{Z}^\mathrm{B}\right) ^{-1} \,\mathbf{U}^{\mathrm{BA}},
\end{equation}
whereby the triple superscript $^{\mathrm{ABA}}$ is used for the sake of clarity. The first superscript defines the modal basis of the scattered field (A), the second superscript describes the scattering object whose scattering properties are to be investigated (B) and the third superscript denotes the modal basis of the incident field (A).\annotateChangesEnd

It connects the incident field coefficients $\mathbf{a}^{\mathrm{A}}$ and the scattered field coefficients $\mathbf{f}^{\mathrm{A}}$:
\begin{equation}
    \mathbf{f}^{\mathrm{A}} = \mathbf{P}^{\mathrm{ABA}} \mathbf{a}^{\mathrm{A}},
\end{equation}
whereby the scattered field is described by:
\begin{equation}
    \mathbf{E}_\mathrm{sc}(\mathbf{r}) = \sum_n f_n^\mathrm{A} \, \mathbf{E}_n^\mathrm{A}(\mathbf{r}).
\end{equation}

It is noted that the perturbation matrix $\mathbf{P}^{\mathrm{ABA}}$ is only a representation of the scattering properties of structure B and does not contain any scattering information about structure A. Since the fields used to describe the incident and scattered field do not diagonalize the scattering operator of structure B in the general case, this perturbation matrix $\mathbf{P}^{\mathrm{ABA}}$ is usually not diagonal, \annotateChangesStart as opposed to \eqref{eq:p_diag}. However, if the structures A and B are equal, the definitions \eqref{eq:p_diag} and \eqref{eq:p_aba_mom_definition} become equal,\annotateChangesEnd{} as shown in Appendix~\ref{sec:coherence_with_existing_cm_def}.

\annotateChangesStart In summary, this novel form of a perturbation matrix formulation enables the scattering properties of structures to be expressed in terms of the CMs of other structures, provided that the surfaces of the latter are a superset of the former. This allows the use of a constant modal basis, when an initial antenna shape is gradually modified during the design process or when two different antennas are compared. \annotateChangesEnd

\section{Transformation Between Characteristic Modes of Different Structures}
\label{sec:transformation}

In the last section, it was shown that the scattering of a structure can be analyzed in terms of the CMs of another structure. While this formulation is sufficient in some cases, it is sometimes also interesting to be able to transform between the CMs of objects. Therefore, it is shown in the following, how the modes of two structures can be transformed into each other.

For this, a modal transformation matrix $\mathbf{Q}^\mathrm{AB}$ is defined. This matrix can be used to transform between the representation of the scattered field 
\begin{equation}
\label{eq:esc_a}
    \mathbf{E}_{\mathrm{sc}}^\mathrm{A}(\mathbf{r}) =  \sum_n f_n^{\mathrm{A}} \mathbf{E}^{\mathrm{A}}_n(\mathbf{r}) 
\end{equation}
and
\begin{equation}
\label{eq:esc_b}
    \mathbf{E}_{\mathrm{sc}}^\mathrm{B}(\mathbf{r}) =  \sum_n f_n^{\mathrm{B}} \mathbf{E}^{\mathrm{B}}_n(\mathbf{r}).
\end{equation}

If $\Omega_\mathrm{B}$ is a subset of $\Omega_\mathrm{A}$
\begin{equation}
\label{eq:omega_b_is_subset_of_omega_a}
    \Omega_\mathrm{B} \subset \Omega_\mathrm{A},
\end{equation}
then an $\mathbf{f}^\mathrm{A}$ can be found for every scattered field $\mathbf{E}_{\mathrm{sc}}^\mathrm{B}(\mathbf{r})$ (for any $\mathbf{f}^{\mathrm{B}}$) radiated by currents on $\Omega_\mathrm{B}$ such that 
\begin{equation}
    \mathbf{E}_{\mathrm{sc}}^\mathrm{A}(\mathbf{r}) = \mathbf{E}_{\mathrm{sc}}^\mathrm{B}(\mathbf{r}) 
\end{equation}
according to:
\begin{equation}
\label{eq:fa_qab_fb}
    \mathbf{f}^{\mathrm{A}} = \mathbf{Q}^\mathrm{AB}  \mathbf{f}^{\mathrm{B}}.
\end{equation}
This is possible since, as mentioned earlier, the fields $\mathbf{E}^{\mathrm{A}}_n(\mathbf{r})$ form the Hilbert space of all fields throughout space produced by currents on $\Omega_\mathrm{A}$, which also includes all possible currents on $\Omega_\mathrm{B}$.

In the other direction, such a statement is strictly not possible if $\Omega_\mathrm{A} \neq \Omega_\mathrm{B}$, since there is not a representation using $\mathbf{E}^{\mathrm{B}}_n(\mathbf{r})$ for every $\mathbf{E}_{\mathrm{sc}}^\mathrm{A}(\mathbf{r})$ produced by currents on $\Omega_\mathrm{A}$. 


For the incident field representations on the surface, the situation is inverted. Since the equation
\begin{equation}
    \left. \mathbf{E}_{\mathrm{inc}}(\mathbf{r})\right|_\mathrm{tan} =  \sum_n a_n^{\mathrm{A}} \mathbf{E}^{0\mathrm{A}}_n(\mathbf{r})  \hspace{3mm}\mathrm{for} \hspace{3mm} \mathbf{r} \in \Omega_\mathrm{A}
\end{equation}
defines the tangential component of $\mathbf{E}_{\mathrm{inc}}(\mathbf{r})$ on a larger set $\Omega_\mathrm{A}$ than the equation
\begin{equation}
    \left. \mathbf{E}_{\mathrm{inc}}(\mathbf{r})\right|_\mathrm{tan} =  \sum_n a_n^{\mathrm{B}} \mathbf{E}^{0\mathrm{B}}_n(\mathbf{r})  \hspace{3mm}\mathrm{for} \hspace{3mm} \mathbf{r} \in \Omega_\mathrm{B},
\end{equation}
there exists a unique mapping from every $\mathbf{a}^{\mathrm{A}}$ to every $\mathbf{a}^{\mathrm{B}}$:
\begin{equation}
\label{eq:ab_eq_qba_aa}
    \mathbf{a}^{\mathrm{B}} = \mathbf{Q}^\mathrm{BA}  \mathbf{a}^{\mathrm{A}}.
\end{equation}
On the other hand, a mapping from $\mathbf{a}^{\mathrm{B}}$ to $\mathbf{a}^{\mathrm{A}}$ would contain the additional degrees of freedom how the tangential component of $\mathbf{E}_{\mathrm{inc}}(\mathbf{r})$ is defined on $\Omega_\mathrm{A} \setminus \Omega_\mathrm{B}$ and would therefore be non-unique.

\annotateChangesStart Assuming, $\mathbf{P}^\mathrm{ABA}$ is the perturbation matrix of scattering object B in the basis of the CMs of structure A and $\mathbf{P}^\mathrm{BBB}$ is the perturbation matrix of scattering object B in the basis of the CMs of structure B, the following transformation rule can be applied between the two:\annotateChangesEnd
\begin{equation}
\label{eq:paba_transformation}
    \mathbf{P}^\mathrm{ABA} = \mathbf{Q}^\mathrm{AB} \mathbf{P}^\mathrm{BBB} \mathbf{Q}^\mathrm{BA}.
\end{equation}


In an EFIE scheme of the method of moments, the transformation matrix \annotateChangesStart $\mathbf{Q}^\mathrm{BA}$ can be calculated according to:
\begin{equation}
\label{eq:q_ab_definition_mom}
    \mathbf{Q}^\mathrm{BA} = \left(\mathbf{I}_{\mathrm{CM}}^\mathrm{B}\right)^\mathrm{T} \,\mathbf{U}^{\mathrm{BA}} = \left(\mathbf{I}_{\mathrm{CM}}^\mathrm{B}\right)^\mathrm{T} \mathbf{R}^{\mathrm{BA}} \mathbf{I}_{\mathrm{CM}}^\mathrm{A},
\end{equation}
which can be derived by decomposing \eqref{eq:vb} using \mbox{$\mathbf{a}^\mathrm{B} = \left(\mathbf{I}_{\mathrm{CM}}^\mathrm{B}\right)^\mathrm{T} \mathbf{V}^\mathrm{B}$} and comparing it to \eqref{eq:ab_eq_qba_aa}.

\annotateChangesEnd Since the characteristic currents are chosen to be real-valued, the transformation matrices are also real-valued. Furthermore, \eqref{eq:q_ab_definition_mom} shows that:
\begin{equation}
    \mathbf{Q}^\mathrm{AB} = \left(\mathbf{Q}^\mathrm{BA}\right)^{\mathrm{T}}.
\end{equation}

Another question is if the transformation is reversible without loss of information in general, which would mean that $\mathbf{Q}^\mathrm{AB}$ would be an orthogonal matrix. \annotateChangesStart It is generally known, that only a finite amount of characteristic modes is calculated correctly, when \eqref{eq:z0_cm_definition} is used to calculate characteristic modes \cite{capek_validating_2017}. Due to this fact, one has to accept that practically: \annotateChangesEnd
\begin{equation}
    \mathbf{Q}^\mathrm{BA} \mathbf{Q}^\mathrm{AB} \neq \mathbf{I}
\end{equation}
and
\begin{equation}
    \mathbf{Q}^\mathrm{AB} \mathbf{Q}^\mathrm{BA} \neq \mathbf{I}.
\end{equation}
In practice, it is observed that for small mode orders the columns of the transformation matrix are indeed approximately orthonormal. \annotateChangesStart For larger mode indices, this property is lost due to the aforementioned numerical inaccuracies in the higher order modes. However, within all investigations throughout our paper, we did not find this to be a limiting factor since the amount of modes calculated correctly according to \eqref{eq:z0_cm_definition} was sufficient for the fields to converge. If additional accuracy is needed in future applications, we suggest to calculate the characteristic modes based on methods shown in \cite{tayli_accurate_2018}.\annotateChangesEnd

\annotateChangesStart Due to missing orthogonality of $\mathbf{Q}^\mathrm{AB}$, \annotateChangesEnd some care has to be taken when scattering matrices are transformed. While a perturbation matrix can be transformed by multiplying with $\mathbf{Q}^\mathrm{AB}$ from the left and $\mathbf{Q}^\mathrm{BA}$ from the right, as seen in \eqref{eq:paba_transformation}, this is not advisable to transform a scattering matrix. Instead, the following formula in terms of scattering matrices should be used:
\begin{equation}
\label{eq:saba_transformation}
\mathbf{S}^\mathrm{ABA} = \mathbf{I} + 2\, \mathbf{P}^\mathrm{ABA} = \mathbf{I} + \mathbf{Q}^\mathrm{AB} \left(\mathbf{S}^\mathrm{BBB} - \mathbf{I} \right) \mathbf{Q}^\mathrm{BA},
\end{equation}
since the scattering matrix $\mathbf{S}^\mathrm{BBB}$ is related to $\mathbf{P}^\mathrm{BBB}$ according to:
\begin{equation*}
    \mathbf{S}^\mathrm{BBB} = \mathbf{I} + 2\, \mathbf{P}^\mathrm{BBB}.
\end{equation*}
It was found that \eqref{eq:saba_transformation} is favorable because otherwise spurious higher order mode currents are introduced.









\section{Examples}
\label{sec:examples}

Now, where the theoretical basis has been derived, \annotateChangesStart three \annotateChangesEnd examples are presented to illustrate the validity of the theory and show its application. Thereby, different aspects like observed convergence in the near-field, in the far-field and the compatibility with non-free-space background media such as the grounded dielectric slab are observed.

\annotateChangesStart \subsection{Two Arbitrary Planar Structures}

In this first example, for validation purposes, the proposed theory is applied for the transformation of the CMs of two arbitrarily chosen structures in a free-space environment, which do not follow any common symmetry and both have multiple significant CMs.

\begin{figure}
    \centering
    \subfloat[][Structure A]{
%
%
\begin{tikzpicture}

\begin{axis}[%
width=1.01in,
height=1.01in,
at={(0.344in,0.326in)},
scale only axis,
xmin=0,
xmax=2,
xlabel style={font=\color{white!15!black}},
xlabel={$x / \lambda_0$},
ymin=0,
ymax=2,
ylabel style={font=\color{white!15!black}},
ylabel={$y / \lambda_0$},
axis background/.style={fill=white},
axis x line*=bottom,
axis y line*=left
]

\addplot[area legend, draw=black, fill=gray, forget plot]
table[row sep=crcr] {%
x	y\\
1.6169153074706	0.273579593182212\\
2	0.449543646123976\\
1.62686555436529	1.71234271384724\\
1	1.55708062233101\\
0	0.811821984384477\\
0.174129507740973	0.32015825805387\\
0.870646790369068	0.568577679313417\\
}--cycle;

\addplot[area legend, draw=black, fill=white, forget plot]
table[row sep=crcr] {%
x	y\\
1.60199012421252	0.651384402511864\\
1.41791036957686	0.661735009056734\\
1.41791036957686	1.199977400259\\
1.62686555436529	1.199977400259\\
}--cycle;
\end{axis}
\end{tikzpicture}%
    }
    \subfloat[][Structure B]{
%
%
\begin{tikzpicture}

\begin{axis}[%
width=1.01in,
height=1.01in,
at={(0.344in,0.326in)},
scale only axis,
xmin=0,
xmax=2,
xlabel style={font=\color{white!15!black}},
xlabel={$x / \lambda_0$},
ymin=0,
ymax=2,
ylabel style={font=\color{white!15!black}},
ylabel={$y / \lambda_0$},
axis background/.style={fill=white},
axis x line*=bottom,
axis y line*=left
]

\addplot[area legend, draw=black, fill=gray, forget plot]
table[row sep=crcr] {%
x	y\\
0.22920851890502	0.650538408895917\\
0.907176802189032	0.722010838147315\\
1.18828540180207	0.62021659082907\\
1.37017926758131	0.85845827111233\\
1.20647482579678	1.2331477585285\\
0.693864806403659	1.26996700213313\\
0.683943370437069	1.13351919500022\\
1.05765253796214	1.11186086042901\\
1.11222080994628	0.98407691095964\\
0.902216084205737	0.834634252751163\\
0.224247800921725	0.760995765541906\\
}--cycle;

\addplot[area legend, dashed, draw=gray, forget plot]
table[row sep=crcr] {%
x	y\\
1.6169153074706	0.273579593182212\\
2	0.449543646123976\\
1.62686555436529	1.71234271384724\\
1	1.55708062233101\\
0	0.811821984384477\\
0.174129507740973	0.32015825805387\\
0.870646790369068	0.568577679313417\\
}--cycle;

\addplot[area legend, dashed, draw=gray, forget plot]
table[row sep=crcr] {%
x	y\\
1.60199012421252	0.651384402511864\\
1.41791036957686	0.661735009056734\\
1.41791036957686	1.199977400259\\
1.62686555436529	1.199977400259\\
}--cycle;
\end{axis}
\end{tikzpicture}%
    }
    \caption{Two arbitrary planar structures in the $xy$-plane at $z = 0$, which are used to illustrate the validity of the proposed theory.}
    \label{fig:arbitrary_structures}
\end{figure}

\begin{figure}
    \centering
    \subfloat[][Eigenvalues of structure A]{
%
%
\definecolor{mycolor1}{rgb}{0.00000,0.44700,0.74100}%
\begin{tikzpicture}

\begin{axis}[%
width=0.96in,
height=1.076in,
at={(0.394in,0.308in)},
scale only axis,
xmin=0.25,
xmax=30.75,
xlabel style={font=\color{white!15!black}},
xlabel={$n$},
ymode=log,
ymin=0.01,
ymax=100,
ytick={0.01,  0.1,    1,   10,  100},
yminorticks=true,
ylabel style={font=\color{white!15!black}},
ylabel={$|\lambda_n^\mathrm{A}|$},
axis background/.style={fill=white}
]
\addplot [color=mycolor1, only marks, mark=x, mark options={solid, mycolor1}, forget plot]
  table[row sep=crcr]{%
1	0.0186475745744076\\
2	0.0747978724324927\\
3	0.153427113452686\\
4	0.168669786956173\\
5	0.184471060581543\\
6	0.217422854689254\\
7	0.224132928648586\\
8	0.411094207878044\\
9	0.426094738638756\\
10	0.590896151024992\\
11	0.903159997693901\\
12	1.06170688535724\\
13	1.09824892817646\\
14	1.50003446352999\\
15	2.67074523289507\\
16	2.88393958106781\\
17	3.68028086433443\\
18	3.87691032867983\\
19	5.00689622937967\\
20	7.35563987641552\\
21	8.59446478218939\\
22	9.39998879665652\\
23	9.84299778196622\\
24	11.8528388265664\\
25	18.3957572161672\\
26	23.5444253922361\\
27	28.2980147372362\\
28	31.183610751958\\
29	36.1752031661387\\
30	47.7684295574986\\
31	55.877181400595\\
32	79.3185788741301\\
33	100.373671924621\\
34	117.54263865076\\
35	167.762273390963\\
36	268.673461727856\\
37	269.036179273547\\
38	370.241414655043\\
39	418.091001927169\\
40	458.274309934823\\
41	513.221333278654\\
42	592.915685978134\\
43	674.171987991185\\
44	912.486864427341\\
45	1124.32362202967\\
46	1844.75553205343\\
47	2452.05283202352\\
48	3278.28157403235\\
49	3837.95942685454\\
};
\end{axis}

\begin{axis}[%
width=1.496in,
height=1.496in,
at={(0in,0in)},
scale only axis,
xmin=0,
xmax=1,
ymin=0,
ymax=1,
axis line style={draw=none},
ticks=none,
axis x line*=bottom,
axis y line*=left
]
\end{axis}
\end{tikzpicture}%
    }
    \subfloat[][Eigenvalues of structure B]{
%
%
\definecolor{mycolor1}{rgb}{0.00000,0.44700,0.74100}%
\begin{tikzpicture}

\begin{axis}[%
width=0.96in,
height=1.076in,
at={(0.394in,0.308in)},
scale only axis,
xmin=0.25,
xmax=30.75,
xlabel style={font=\color{white!15!black}},
xlabel={$n$},
ymode=log,
ymin=0.01,
ymax=100,
ytick={0.01,  0.1,    1,   10,  100},
yminorticks=true,
ylabel style={font=\color{white!15!black}},
ylabel={$|\lambda_n^\mathrm{B}|$},
axis background/.style={fill=white}
]
\addplot [color=mycolor1, only marks, mark=x, mark options={solid, mycolor1}, forget plot]
  table[row sep=crcr]{%
1	0.241944131348475\\
2	0.460366070861342\\
3	1.20994458574727\\
4	1.71289776205127\\
5	3.85747409983023\\
6	9.54723998296745\\
7	14.8549385620147\\
8	23.2664861729642\\
9	24.1697711458659\\
10	29.5945616537636\\
11	48.508640475847\\
12	52.5445229840578\\
13	109.591412600417\\
14	215.69349616852\\
15	263.992620160433\\
16	372.793283406466\\
17	433.876546391955\\
18	795.483342686729\\
19	1256.50376314094\\
20	2748.18207388832\\
21	2907.86149674825\\
22	6362.85988768281\\
23	6707.62136859847\\
24	13096.3338268426\\
25	18239.9890802925\\
26	29966.6026919243\\
27	32040.6866486515\\
28	47026.8358220998\\
29	57224.2466972026\\
30	137301.786057537\\
31	138576.903451705\\
32	197659.257674061\\
33	247666.063288968\\
34	259367.590956785\\
35	287315.873695706\\
36	372705.588540184\\
37	455907.575619627\\
38	460499.151205673\\
39	582538.172587839\\
40	759815.718569414\\
41	980145.588245754\\
42	1044339.54965754\\
43	1123886.68077839\\
44	1583638.60063419\\
45	2368497.59112659\\
46	2426160.7932777\\
};
\end{axis}
\end{tikzpicture}%
    }
    \caption{Eigenvalues of the arbitrary structures that are shown in Fig.\ref{fig:arbitrary_structures}.}
    \label{fig:arbitrary_structure_eigenvalues}
\end{figure}

The structures are shown in Fig.~\ref{fig:arbitrary_structures}. Both structures are polygons at $z = 0$, whereby structure B is a substructure of structure A, as it is required for the proposed theory to converge. To validate the theory with more than two modes, both structures are selected to have an electrical size of at least $1\lambda_0$.

Now, both structures are meshed independently. Their impedance matrices $\mathbf{Z}^\mathrm{A}$ and $\mathbf{Z}^\mathrm{B}$ are assembled using our in-house method of moments code \cite{noauthor_-house_nodate} and the eigencurrents $\mathbf{I}^{\mathrm{A}}_\mathrm{CM}$ and $\mathbf{I}^{\mathrm{B}}_\mathrm{CM}$ and eigenvalues $\lambda_n^\mathrm{A}$ and $\lambda_n^\mathrm{B}$ are calculated individually according to \eqref{eq:z0_cm_definition}. The magnitude of the eigenvalues of both structures are shown in Fig.~\ref{fig:arbitrary_structure_eigenvalues}. It is seen that, as it was desired, both structures have multiple significant modes when a significance threshold of $|\lambda_n| < 1$ is chosen to cover all modes that could be well-excited\cite{manteuffel_compact_2016}.  

Now, in order to show the validity of the proposed theory, the goal is to show that the characteristic near-fields of the two structures can be transformed into each other. For the sake of simplicity, but without loss of generality, the transformation formulas \eqref{eq:esc_a}, \eqref{eq:esc_b} and \eqref{eq:fa_qab_fb} are reduced to the case of a single $\nu$\nobreakdash-th CM of structure B:
\begin{equation}
\label{eq:E_nu_exact}
    \mathbf{E}_{\nu}^{\mathrm{B}}(\mathbf{r}) = \sum_{n = 1}^\infty \mathbf{E}_n^\mathrm{A}(\mathbf{r}) q_{\nu n}^{\mathrm{AB}},
\end{equation}
whereby $q_{\nu n}^{\mathrm{AB}}$ are the matrix entries of $\mathbf{Q}^{\mathrm{AB}}$.

By truncating the sum in \eqref{eq:E_nu_exact} after the $N_\mathrm{A}$\nobreakdash-th CMs of structure A, an approximation of $\mathbf{E}_{\nu}^{\mathrm{B}}(\mathbf{r})$ can be defined:
\begin{equation}
    \mathbf{E}_{\nu}^{\mathrm{B}}(\mathbf{r}) \approx \hat{\mathbf{E}}_{\nu,N_\mathrm{A}}^\mathrm{B}(\mathbf{r}) = \sum_{n = 1}^{N_\mathrm{A}} \mathbf{E}_n^\mathrm{A}(\mathbf{r}) \, q_{\nu n}^{\mathrm{AB}}.
\end{equation}

In the following, it will be shown that this approximation converges to the value of $\mathbf{E}_{\nu}^{\mathrm{B}}(\mathbf{r})$, exemplary for the first mode of structure B.
For the observation points, a path $0.4\lambda_0$ above the structure is chosen arbitrarily. The near-fields $\mathbf{E}_n^\mathrm{A}(x, \lambda_0, 0.4\lambda_0)$ and $\mathbf{E}_n^\mathrm{B}(x, \lambda_0, 0.4\lambda_0)$ are calculated using the in-house method of moments code. Furthermore, the modal transformation matrix $\mathbf{Q}^{\mathrm{AB}}$ is calculated using \eqref{eq:q_ab_definition_mom}.

\begin{figure}
    \centering
    \input{images/arbitrary_structure_e_field}
    \caption{Magnitude of the first characteristic field of structure B along the chosen observation path from a direct calculation of the near-field $\mathbf{E}_1^{\mathrm{B}}(x,\lambda_0, 0.4\lambda_0)$ (\lineleg{mycolor1}) and by superposition of the characteristic fields of structure A $\hat{\mathbf{E}}_{1,N_\mathrm{A}}^\mathrm{B}(x,\lambda_0, 0.4\lambda_0)$ for \mbox{$N_\mathrm{A} = 1$} (\lineleg{mycolor2,dashed}), for \mbox{$N_\mathrm{A} = 5$} (\lineleg{mycolor3,dashed}), for \mbox{$N_\mathrm{A} = 10$} (\lineleg{mycolor4,dashed}), for \mbox{$N_\mathrm{A} = 20$} (\lineleg{mycolor5,dashed}) and for \mbox{$N_\mathrm{A} = 30$} (\lineleg{mycolor6,dashed}) modes.}
    \label{fig:arbitrary_structure_e_field}
\end{figure}

Fig.~\ref{fig:arbitrary_structure_e_field} shows the magnitude of the first characteristic field of structure B $\mathbf{E}_1^\mathrm{B}(x, \lambda_0, 0.4\lambda_0)$ and its approximation by a superposition of the characteristic fields of structure A $\hat{\mathbf{E}}_{\nu,N_\mathrm{A}}^\mathrm{B}(\mathbf{r})$ for different number of used modes $N_\mathrm{A}$. 

It is seen that the approximated value approaches the directly calculated value for large $N_\mathrm{A}$. Since there was no common symmetry or other similarity between both structures, this can be seen as a general indication of the validity of the proposed theory.

\annotateChangesEnd

\subsection{Dipole with Varying Length}

\begin{figure}
    \centering
    \includegraphics{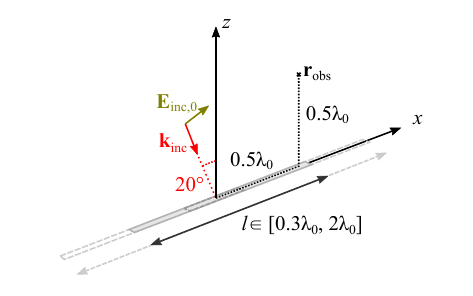}
    \caption{Exemplary scattering scenario of a dipole with varying length $l$.}
    \label{fig:dipole_scattering_1}
\end{figure}

In this example, \annotateChangesStart  some properties of the perturbation matrices are shown and the validity of the proposed method is illustrated further in a scattering scenario. For that purpose, \annotateChangesEnd the effects of changing the length of a dipole $l$ are investigated in a simple scattering scenario, as shown in Fig.~\ref{fig:dipole_scattering_1}.


\annotateChangesStart
First, the perturbation matrices are calculated and compared for two exemplary dipole lengths \mbox{$l = 2\lambda_0$ and $l = \lambda_0$}. \annotateChangesEnd In order to calculate the length-dependent perturbation matrices \mbox{$\mathbf{P}(l)$}, an impedance-matrix $\mathbf{Z}$ is assembled for both dipole lengths using our in-house method of moments code \cite{noauthor_-house_nodate}. Then, the CMs of each structure are at first calculated in the eigenbasis of the structure according to \eqref{eq:z0_cm_definition}, leading to the diagonal perturbation matrices:
\begin{equation}
    \mathbf{P}^\prime(l) = - \operatorname{diag} \frac{1}{1+\mathrm{j} \lambda_n(l)}.
\end{equation}

Furthermore, the modal transformation matrices \mbox{$\mathbf{Q}(l)$} are calculated according to \eqref{eq:q_ab_definition_mom} to transform between the modes of the reference dipole (of length $2\lambda_0$) and the modes of the dipole with length $l$. By combining the former two, the perturbation matrices \mbox{$\mathbf{P}(l)$} in the reference basis are then calculated using:
\begin{equation}
\label{eq:P_of_l}
    \mathbf{P}(l) = \mathbf{Q}(l)^\mathrm{T} \mathbf{P}^\prime(l) \mathbf{Q}(l).
\end{equation}


The perturbation matrices, in the reference basis \mbox{$\mathbf{P}(l)$} and in the eigenbasis \mbox{$\mathbf{P}^\prime(l)$}, are exemplary shown in Fig.~\ref{fig:dipole_perturbation_matrices} for the two different dipole lengths.

When the perturbation matrices \mbox{$\mathbf{P}^\prime(2\lambda_0)$} and \mbox{$\mathbf{P}(2\lambda_0)$} are compared, it is found that both matrices are equal and diagonal. This is since \mbox{$l = 2\lambda_0$} is the reference length for the modes used for $\mathbf{P}(l)$, which means that the two decompositions are equal and \mbox{$\mathbf{Q}(2\lambda_0) = \mathbf{I}$}. The diagonality of the perturbation matrices means that each incident CM is only scattered into the corresponding outgoing CM with the same mode index.

On the other hand, for a dipole length of \mbox{$l = \lambda_0$}, the perturbation matrix in the reference basis \mbox{$\mathbf{P}(\lambda_0)$} significantly differs from the diagonal perturbation matrix in the eigenbasis \mbox{$\mathbf{P}^\prime(\lambda_0)$}. This is since the CM basis of the dipole with length $2\lambda_0$ is not an eigenbasis of the dipole with length $\lambda_0$, which means that each incident CM is scattered also into multiple outgoing mode indices.

\begin{figure}
    \centering
    \include{images/dipole_perturbation_matrices}
    \caption{First six columns and rows of the perturbation matrices \mbox{$\mathbf{P}(l)$} and \mbox{$\mathbf{P}^\prime(l)$} for the dipole lengths \mbox{$l = 2 \lambda_0$} and \mbox{$l = \lambda_0$} and the two dimensional colormap that is used to encode the phase and the magnitude of the entries. The index $n$ encodes the incident mode index and the index $\nu$ encodes the scattered mode index.}
    \label{fig:dipole_perturbation_matrices}
\end{figure}

\annotateChangesStart
Now, to illustrate the validity of the proposed theory, the scattered near-field at an arbitrarily chosen observation point \mbox{$\mathbf{E}_{\mathrm{sc}}(\mathbf{r}_\mathrm{obs})$} is to be calculated and compared against reference simulations. The eigenfields of the longest dipole (with length $2\lambda_0$) are used to describe the scattering problem for all dipole lengths \mbox{$l \in [0.3\lambda_0, 2\lambda_0]$} as proposed within this paper. Therefore, the perturbation matrix $\mathbf{P}(l)$ is now calculated for additional samples of $l$ within that interval according to \eqref{eq:P_of_l}. Furthermore, the characteristic near-fields \mbox{$\mathbf{E}_n(\mathbf{r}_\mathrm{obs}, 2\lambda_0)$} and the modal weighting coefficients of the incident plane wave $a_n$ are calculated for the reference dipole with length $2\lambda_0$ using the in-house method of moments code according to:
\begin{equation}
    a_n = \int \mathbf{J}_n\, \mathbf{E}_{\mathrm{inc}}\, \mathrm{d}S = \mathbf{I}_n^\mathrm{T} \mathbf{V}_{\mathrm{inc}}.
\end{equation}

Since the modal fields and the modal weighting coefficients of the incident field $a_n$ remain constant with respect to a change of the dipole length $l$, all changes of the structure are contained within the length-dependent perturbation matrix $\mathbf{P}(l)$:
\begin{equation}
\label{eq:f_of_l}
    \mathbf{f}(l) = \mathbf{P}(l) \, \mathbf{a},
\end{equation}
whereby the coefficients $f_n(l)$ are the weighting coefficients of the scattered field. \annotateChangesEnd 

Then, \eqref{eq:f_of_l} is used to calculate the scattered field coefficients $f_n(l)$ from $a_n$ and the perturbation matrices \mbox{$\mathbf{P}(l)$}. Consequently, the scattered field is calculated according to:
\begin{equation}
\label{eq:sum_e_sc_for_dipole}
    \mathbf{E}_{\mathrm{sc}}(\mathbf{r}_{\mathrm{obs}}, l) = \sum_{n=1}^{N}\mathbf{E}_n(\mathbf{r}_{\mathrm{obs}}, 2\lambda_0)\, f_n (l).
\end{equation}

\begin{figure}
    \centering
    \input{images/dipole_scattered_field}
    \caption{Scattered field at the chosen observation point \mbox{$\mathbf{r}_\mathrm{obs} = [0.5\lambda_0, 0, 0.5\lambda_0]
    $} for different electrical lengths of the dipole $l/\lambda_0$ from a direct method of moments solution (\lineleg{mycolor1}), by the characteristic fields and the perturbation matrix $\mathbf{P}(l)$ according to \eqref{eq:P_of_l} with \mbox{$N = 1$} mode (\lineleg{mycolor3,dashed}), \mbox{$N = 2$} modes (\lineleg{mycolor4,dashed}), \mbox{$N = 4$} modes (\lineleg{mycolor5,dashed}), \mbox{$ N = 6$} modes (\lineleg{mycolor6,dashed}) and by the characteristic fields and the approximated perturbation matrix \mbox{$\mathbf{P}_{\mathrm{approx}}(l)$} according to \eqref{eq:P_l_approx_p_prime_l} (\lineleg{mycolor2}). }
    \label{fig:dipole_scattered_field}
\end{figure}

Now, in Fig.~\ref{fig:dipole_scattered_field}, the scattered field is shown for all lengths~$l$ for different number of modes $N$ to be used within \eqref{eq:sum_e_sc_for_dipole}. For reference, a full-wave simulation is also conducted for each dipole length $l$ directly using the method of moments. Good agreement between both curves is found for \mbox{$N = 4$} and \mbox{$N = 6$} modes. \annotateChangesStart This indicates a good convergence in the near-field of the dipole and illustrates the validity of the approach proposed within this paper. In addition, \annotateChangesEnd the convergence is also shown by the decay of the normalized error magnitude of the electric field for an arbitrarily chosen length of \mbox{$l = 0.8\lambda_0$} in Fig.~\ref{fig:dipole_scattered_field_error}.

\begin{figure}
    \centering
%
%
\definecolor{mycolor1}{rgb}{0.00000,0.44700,0.74100}%
\begin{tikzpicture}

\begin{axis}[%
width=2.422in,
height=1.446in,
at={(0.428in,0.375in)},
scale only axis,
xmin=0.272727272727273,
xmax=11.7272727272727,
xlabel style={font=\color{white!15!black}},
xlabel={$N$},
ymin=0,
ymax=1,
ylabel style={font=\color{white!15!black}},
ylabel={$|\sum_{n=1}^N \mathbf{E}_n f_n - \mathbf{E}_{\mathrm{sc}}|/|\mathbf{E}_{\mathrm{sc}}|$},
axis background/.style={fill=white}
]
\addplot[ycomb, color=mycolor1, mark=o, mark options={solid, mycolor1}, forget plot] table[row sep=crcr] {%
1	0.919381863996518\\
2	0.906489801405927\\
3	0.486589175724555\\
4	0.0708933416895261\\
5	0.0721069301794587\\
6	0.0172132540762986\\
7	0.0166446969371267\\
8	0.0166323909241317\\
9	0.0166319719636078\\
10	0.0166328878692658\\
11	0.0166134231099297\\
};
\addplot[forget plot, color=white!15!black] table[row sep=crcr] {%
0.272727272727273	0\\
11.7272727272727	0\\
};
\end{axis}
\end{tikzpicture}%
    \caption{Normalized error magnitude of the electric field as a function of the number of CMs of the base structure $N$ used to describe the scattered field for a dipole length \mbox{$l = 0.8\lambda_0$} at the chosen observation point \mbox{$\mathbf{r}_\mathrm{obs} = [0.5\lambda_0, 0, 0.5\lambda_0]
    $}.}
    \label{fig:dipole_scattered_field_error}
\end{figure}

Furthermore, as discussed in the introduction, intuitive design methods based on characteristic modes often implicitly assume that the eigenfields do not change significantly with geometric changes. Therefore, for comparison, Fig.~\ref{fig:dipole_scattered_field} also shows how this approximation
\begin{equation}
    \mathbf{E}_{n}(\mathbf{r}_{\mathrm{obs}}, l) \approx \mathbf{E}_{n}(\mathbf{r}_{\mathrm{obs}}, 2\lambda_0),
\end{equation}
which can also be written as
\begin{equation}
\label{eq:P_l_approx_p_prime_l}
    \mathbf{P}_{\mathrm{approx}}(l) \approx \mathbf{P}^\prime(l),
\end{equation}
performs in contrast to the proposed formalism. Here, significant disagreement in comparison to the direct method of moments solution is observed for \mbox{$l < 1.8\lambda_0$}, meaning that this approximation is only valid close to the reference length of $2\lambda_0$.

In summary, it was seen that a dipole with varying length was represented in a fixed CM basis. It was thereby found that the scattered near-field at an arbitrarily chosen point in front of the dipole converges well in that basis. This illustrates the validity of the proposed approach.

\subsection{Patch Antenna with a Radiation Null}

In this example, another geometry variation of a structure is investigated. While the former \annotateChangesStart examples were \annotateChangesEnd focused on computational aspects and the illustration of the perturbation matrices, the focus here is on the aspect that the proposed formalism can also be used to obtain a more intuitive decomposition than the eigendecomposition of the structure in certain cases.

\begin{figure}
    \centering
    \includegraphics{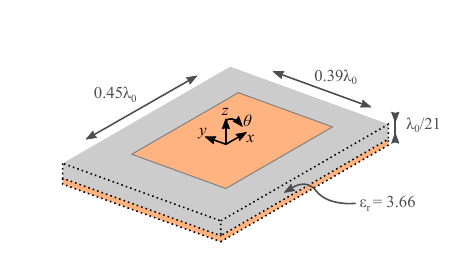}
    \caption{Geometry of the initial base structure of the patch antenna.}
    \label{fig:patch_scattering_base2}
\end{figure}

The chosen generic example is the design of a $\theta$-polarized, \annotateChangesStart probe-fed \annotateChangesEnd patch antenna that is intended to have a radiation null at a pre-defined direction. \annotateChangesStart The goal is to create a tilted beam that has a radiation null at \mbox{$\theta = -15^\circ$} in the $xz$\nobreakdash-plane. \annotateChangesEnd An initial structure is proposed, as shown in Fig.~\ref{fig:patch_scattering_base2}. It consists of a patch on an infinitely extended grounded dielectric slab \cite{mrochen_characteristic_2025}. 

\begin{figure}
    \centering
    
    \subfloat[][$\mathbf{J}_1$ with $\lambda_1 = -3.1$.]{
        \begin{tikzpicture}[scale=0.3]
            \input{images/patch_initial_eigencurrents-1}
            \draw [ultra thick, white, arrows = {-Stealth}] (15mm,0) -- (35mm,0);
            \draw [ultra thick, white, arrows = {-Stealth}] (-15mm,0) -- (-35mm,0);
        \end{tikzpicture}
    }\hspace{4mm}
    \subfloat[][$\mathbf{J}_2$ with $\lambda_2 = 3.9$.]{
        \begin{tikzpicture}[scale=0.3]
            \input{images/patch_initial_eigencurrents-2}
            \draw [ultra thick, white, arrows = {-Stealth}] (0,-15mm) -- (0,15mm);
        \end{tikzpicture}
    } \\
    \vspace{4mm}
    \subfloat[][$\mathbf{J}_3$ with $\lambda_3 = 4.3$.]{
        \begin{tikzpicture}[scale=0.3]
            \input{images/patch_initial_eigencurrents-3}
            \draw [ultra thick, white, arrows = {-Stealth}] (-20mm,0) -- (20mm, 0);
        \end{tikzpicture}
    }\hspace{4mm}
    \subfloat[][$\mathbf{J}_4$ with $\lambda_4 = 7.0$.]{
        \begin{tikzpicture}[scale=0.3]
            \input{images/patch_initial_eigencurrents-4}
            \draw [ultra thick, white, arrows = {-Stealth}] (-32mm,-10mm) -- (-32mm,10mm);
            \draw [ultra thick, white, arrows = {-Stealth}] (32mm,10mm) -- (32mm,-10mm);
            \draw [ultra thick, white, arrows = {-Stealth}] (20mm,27mm) -- (-20mm, 27mm);
            \draw [ultra thick, white, arrows = {-Stealth}] (-20mm,-27mm) -- (20mm, -27mm);
        \end{tikzpicture}
    }\vspace{-15mm}
    
    \hspace{-80mm}
    \begin{tikzpicture}[scale=0.3]
        \draw [arrows = {-Stealth}] (0,0) -- (0,15mm);
        \draw [arrows = {-Stealth}] (0,0) -- (15mm,0);
        \node at (20mm,0) {$x$};
        \node at (0,22mm) {$y$};
    \end{tikzpicture}\vspace{4mm}

    \caption{\annotateChangesStart The first four characteristic current distributions $\mathbf{J}_n$ of the initial structure at the target frequency ordered by the magnitude of their eigenvalues $|\lambda_n|$.\annotateChangesEnd}
    \label{fig:patch_initial_currents}
\end{figure}

\begin{figure}
    \centering
%
%
\definecolor{mycolor1}{rgb}{0.00000,0.44700,0.74100}%
\definecolor{mycolor2}{rgb}{0.85000,0.32500,0.09800}%
\definecolor{mycolor3}{rgb}{0.92900,0.69400,0.12500}%
\begin{tikzpicture}

\begin{axis}[%
width=2.307in,
height=1.273in,
at={(0.543in,0.501in)},
scale only axis,
xmin=-90,
xmax=90,
xtick={-90, -60, -30,   0,  30,  60,  90},
xlabel style={font=\color{white!15!black}},
xlabel={$\theta\text{ in }{\circ}$},
ymin=-10,
ymax=15,
ylabel style={font=\color{white!15!black}},
ylabel={Directivity in dBi},
axis background/.style={fill=white},
xmajorgrids,
ymajorgrids
]
\addplot [color=mycolor1, line width=1.0pt, forget plot]
  table[row sep=crcr]{%
-180	-20.5804616645864\\
-175	-20.5804616645864\\
-170	-20.5804616645864\\
-165	-20.5804616645864\\
-160	-20.5804616645864\\
-155	-20.5804616645864\\
-150	-20.5804616645864\\
-145	-20.5804616645864\\
-140	-20.5804616645864\\
-135	-20.5804616645864\\
-130	-20.5804616645864\\
-125	-20.5804616645864\\
-120	-20.5804616645864\\
-115	-20.5804616645864\\
-110	-20.5804616645864\\
-105	-20.5804616645864\\
-100	-20.5804616645864\\
-95	-20.5804616645864\\
-90	-20.5804616645864\\
-85	-20.5804616645864\\
-80	-13.5582012282876\\
-75	-6.75114654338618\\
-70	-2.09095419711052\\
-65	1.34098763323891\\
-60	3.94816296548806\\
-55	5.93927903331796\\
-50	7.43129613998504\\
-45	8.49045958256861\\
-40	9.15038815685999\\
-35	9.41953833541356\\
-30	9.28192083636998\\
-25	8.69075111801327\\
-20	7.54964736528213\\
-15	5.66224454778262\\
-10	2.57277683882174\\
-5	-3.19007117194684\\
0	-20.5804616645864\\
-0	-20.5804616645864\\
5	-3.19007117194684\\
10	2.57277683882174\\
15	5.66224454778262\\
20	7.54964736528214\\
25	8.69075111801326\\
30	9.28192083636998\\
35	9.41953833541356\\
40	9.15038815686\\
45	8.4904595825686\\
50	7.43129613998505\\
55	5.93927903331795\\
60	3.94816296548806\\
65	1.3409876332389\\
70	-2.09095419711052\\
75	-6.75114654338618\\
80	-13.5582012282876\\
85	-20.5804616645864\\
90	-20.5804616645864\\
95	-20.5804616645864\\
100	-20.5804616645864\\
105	-20.5804616645864\\
110	-20.5804616645864\\
115	-20.5804616645864\\
120	-20.5804616645864\\
125	-20.5804616645864\\
130	-20.5804616645864\\
135	-20.5804616645864\\
140	-20.5804616645864\\
145	-20.5804616645864\\
150	-20.5804616645864\\
155	-20.5804616645864\\
160	-20.5804616645864\\
165	-20.5804616645864\\
170	-20.5804616645864\\
175	-20.5804616645864\\
180	-20.5804616645864\\
180	-20.5804616645864\\
};
\addplot [color=mycolor2, line width=1.0pt, forget plot]
  table[row sep=crcr]{%
-180	-20.5248373530783\\
-175	-20.5248373530783\\
-170	-20.5248373530783\\
-165	-20.5248373530783\\
-160	-20.5248373530783\\
-155	-20.5248373530783\\
-150	-20.5248373530783\\
-145	-20.5248373530783\\
-140	-20.5248373530783\\
-135	-20.5248373530783\\
-130	-20.5248373530783\\
-125	-20.5248373530783\\
-120	-20.5248373530783\\
-115	-20.5248373530783\\
-110	-20.5248373530783\\
-105	-20.5248373530783\\
-100	-20.5248373530783\\
-95	-20.5248373530783\\
-90	-20.5248373530783\\
-85	-20.5248373530783\\
-80	-20.5248373530783\\
-75	-15.3452022466899\\
-70	-10.4295249784559\\
-65	-6.66336907157028\\
-60	-3.63826513089993\\
-55	-1.13854214123021\\
-50	0.962414953278412\\
-45	2.74429137342316\\
-40	4.25967127953546\\
-35	5.54440484570368\\
-30	6.62347204436122\\
-25	7.51450532963494\\
-20	8.23000482532928\\
-15	8.7787768879648\\
-10	9.16688444991136\\
-5	9.39827242116408\\
0	9.47516264692172\\
-0	9.47516264692172\\
5	9.39827242116408\\
10	9.16688444991136\\
15	8.77877688796479\\
20	8.23000482532927\\
25	7.51450532963492\\
30	6.6234720443612\\
35	5.54440484570367\\
40	4.25967127953543\\
45	2.74429137342313\\
50	0.96241495327839\\
55	-1.13854214123023\\
60	-3.63826513089995\\
65	-6.66336907157031\\
70	-10.4295249784559\\
75	-15.3452022466899\\
80	-20.5248373530783\\
85	-20.5248373530783\\
90	-20.5248373530783\\
95	-20.5248373530783\\
100	-20.5248373530783\\
105	-20.5248373530783\\
110	-20.5248373530783\\
115	-20.5248373530783\\
120	-20.5248373530783\\
125	-20.5248373530783\\
130	-20.5248373530783\\
135	-20.5248373530783\\
140	-20.5248373530783\\
145	-20.5248373530783\\
150	-20.5248373530783\\
155	-20.5248373530783\\
160	-20.5248373530783\\
165	-20.5248373530783\\
170	-20.5248373530783\\
175	-20.5248373530783\\
180	-20.5248373530783\\
180	-20.5248373530783\\
};
\addplot [color=mycolor3, line width=1.0pt, forget plot]
  table[row sep=crcr]{%
-180	-18.8428323501363\\
-175	-18.8428323501363\\
-170	-18.8428323501363\\
-165	-18.8428323501363\\
-160	-18.8428323501363\\
-155	-18.8428323501363\\
-150	-18.8428323501363\\
-145	-18.8428323501363\\
-140	-18.8428323501363\\
-135	-18.8428323501363\\
-130	-18.8428323501363\\
-125	-18.8428323501363\\
-120	-18.8428323501363\\
-115	-18.8428323501363\\
-110	-18.8428323501363\\
-105	-18.8428323501363\\
-100	-18.8428323501363\\
-95	-18.8428323501363\\
-90	-18.8428323501363\\
-85	-18.8428323501363\\
-80	-17.6799002399037\\
-75	-10.9329883727652\\
-70	-6.36096470099582\\
-65	-3.04979005182923\\
-60	-0.602890563266682\\
-55	1.17804700535748\\
-50	2.39407495121002\\
-45	3.08623419182161\\
-40	3.2460666524434\\
-35	2.80631853709984\\
-30	1.59898388044247\\
-25	-0.785768236586076\\
-20	-5.64724265755635\\
-15	-18.8428323501363\\
-10	-5.95954588655439\\
-5	0.719575683417866\\
0	4.47443425634686\\
-0	4.47443425634686\\
5	6.97337017444207\\
10	8.71533520563145\\
15	9.9162329487277\\
20	10.6869414677406\\
25	11.089142878895\\
30	11.1571676498637\\
35	10.9074929454387\\
40	10.3427118467438\\
45	9.45235196534005\\
50	8.21116660958506\\
55	6.57441355214241\\
60	4.4682399068593\\
65	1.77036681998086\\
70	-1.73167552359198\\
75	-6.44403118709731\\
80	-13.2871559800937\\
85	-18.8428323501363\\
90	-18.8428323501363\\
95	-18.8428323501363\\
100	-18.8428323501363\\
105	-18.8428323501363\\
110	-18.8428323501363\\
115	-18.8428323501363\\
120	-18.8428323501363\\
125	-18.8428323501363\\
130	-18.8428323501363\\
135	-18.8428323501363\\
140	-18.8428323501363\\
145	-18.8428323501363\\
150	-18.8428323501363\\
155	-18.8428323501363\\
160	-18.8428323501363\\
165	-18.8428323501363\\
170	-18.8428323501363\\
175	-18.8428323501363\\
180	-18.8428323501363\\
180	-18.8428323501363\\
};
\end{axis}
\end{tikzpicture}%
    \caption{Directivity of the far-field of the first CM of the base structure (\lineleg{mycolor1}) and the third CM of the base structure (\lineleg{mycolor2}) and the desired far-field as a superposition of the first and third CM of the base structure (\lineleg{mycolor3}).}
    \label{fig:patch_modes2}
\end{figure}

In order to realize \annotateChangesStart the tilted beam \annotateChangesEnd on a single antenna, multiple modes have to be used. \annotateChangesStart The first four CMs current distributions on the base structure are calculated and depicted in Fig.~\ref{fig:patch_initial_currents} along with their eigenvalues. Since only the modes $\mathbf{J}_1$ and $\mathbf{J}_3$ radiate the desired $\theta$\nobreakdash-polarization the $xz$\nobreakdash-plane, only these modes are used in the following. \annotateChangesEnd Fig.~\ref{fig:patch_modes2} shows the far-fields of these CMs. Using single-element beamforming~\cite{morlein_understanding_2022}, a superposition of the two modes is found where a radiation null at $\theta = -15^\circ$ is achieved:
\begin{equation}
    \mathbf{F}(\theta) = f_1 \mathbf{F}_1(\theta) + f_3 \mathbf{F}_3(\theta),
\end{equation}
whereby $\mathbf{F}_1(\theta)$ and $\mathbf{F}_3(\theta)$ are the far-fields of the first and third CM of the base structure. The optimal ratio of the modal weighting coefficients is found to be \mbox{$|f_1/f_3| = 2.6$} in magnitude with a phase difference of \mbox{$\angle f_1/f_3 = 90^\circ$}. The resulting pattern $\mathbf{F}(\theta)$ is also depicted in Fig.~\ref{fig:patch_modes2}.

\begin{figure}
    \centering
    \includegraphics{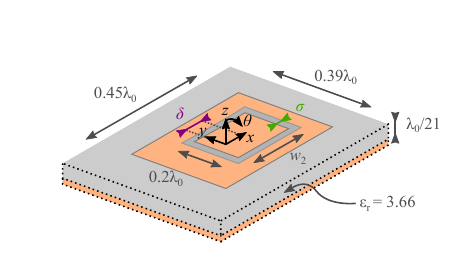}
    \caption{Geometry of the modified structure of the patch antenna.}
    \label{fig:patch_scattering_variation2}
\end{figure}

Now, to actually realize such a pattern, both modes have to be \say{excitable} on the structure. This is usually achieved if the eigenvalues are in the range \mbox{$|\lambda_n| < 1$} \cite{manteuffel_compact_2016}. However, for the base structure, it is found that \mbox{$\lambda_1=-3.1$} and \mbox{$\lambda_3 = 4.3$}, which means that the first mode is too capacitive and the third mode is too inductive in order to be excited properly.

This can be changed by modifying the structure, similar as done in \cite{gausmann_multi-mode_2025}. An inner patch is introduced to support the broadside mode, which makes the mode more capacitive due to the smaller resonant length. On the other hand, the current path on the outer patch is elongated by the detour enforced by the gap, making the off-broadside mode more inductive. The modified structure is shown in Fig.~\ref{fig:patch_scattering_variation2}, whereby the gap is chosen to be \mbox{$\sigma = 0.01 \lambda_0$} and the inner patch is centered ($\delta = 0$). By tuning the width of the inner patch $w_2$, both modes are shifted towards the excitable range, yielding a width of \mbox{$w_2 = 0.235 \lambda_0$}.

Now, to excite the structure, a single probe-feed should be used. Since the excitation of the cross-polarized modes is undesired, the probe-feed is placed on the $x$\nobreakdash-axis. The modal weighting coefficients for the varying probe-feed position $x_\mathrm{P}$ are calculated based on the assumption of an impressed thin current filament, which yields in a computational formula based on a weighted superposition of the eigenfields of the modified structure, as described in \cite{yang_computing_2016}.

\begin{figure}
    \centering
%
%
\definecolor{mycolor1}{rgb}{0.00000,0.44700,0.74100}%
\definecolor{mycolor2}{rgb}{0.85000,0.32500,0.09800}%
\begin{tikzpicture}

\begin{axis}[%
width=2.307in,
height=1.273in,
at={(0.543in,0.501in)},
scale only axis,
xmin=-90,
xmax=90,
xtick={-90, -60, -30,   0,  30,  60,  90},
xlabel style={font=\color{white!15!black}},
xlabel={$\theta\text{ in }{\circ}$},
ymin=-10,
ymax=15,
ylabel style={font=\color{white!15!black}},
ylabel={Directivity in dBi},
axis background/.style={fill=white},
xmajorgrids,
ymajorgrids
]
\addplot [color=mycolor1, dashed, line width=1.0pt, forget plot]
  table[row sep=crcr]{%
-180	-20.5804616645864\\
-175	-20.5804616645864\\
-170	-20.5804616645864\\
-165	-20.5804616645864\\
-160	-20.5804616645864\\
-155	-20.5804616645864\\
-150	-20.5804616645864\\
-145	-20.5804616645864\\
-140	-20.5804616645864\\
-135	-20.5804616645864\\
-130	-20.5804616645864\\
-125	-20.5804616645864\\
-120	-20.5804616645864\\
-115	-20.5804616645864\\
-110	-20.5804616645864\\
-105	-20.5804616645864\\
-100	-20.5804616645864\\
-95	-20.5804616645864\\
-90	-20.5804616645864\\
-85	-20.5804616645864\\
-80	-13.5582012282876\\
-75	-6.75114654338618\\
-70	-2.09095419711052\\
-65	1.34098763323891\\
-60	3.94816296548806\\
-55	5.93927903331796\\
-50	7.43129613998504\\
-45	8.49045958256861\\
-40	9.15038815685999\\
-35	9.41953833541356\\
-30	9.28192083636998\\
-25	8.69075111801327\\
-20	7.54964736528213\\
-15	5.66224454778262\\
-10	2.57277683882174\\
-5	-3.19007117194684\\
0	-20.5804616645864\\
-0	-20.5804616645864\\
5	-3.19007117194684\\
10	2.57277683882174\\
15	5.66224454778262\\
20	7.54964736528214\\
25	8.69075111801326\\
30	9.28192083636998\\
35	9.41953833541356\\
40	9.15038815686\\
45	8.4904595825686\\
50	7.43129613998505\\
55	5.93927903331795\\
60	3.94816296548806\\
65	1.3409876332389\\
70	-2.09095419711052\\
75	-6.75114654338618\\
80	-13.5582012282876\\
85	-20.5804616645864\\
90	-20.5804616645864\\
95	-20.5804616645864\\
100	-20.5804616645864\\
105	-20.5804616645864\\
110	-20.5804616645864\\
115	-20.5804616645864\\
120	-20.5804616645864\\
125	-20.5804616645864\\
130	-20.5804616645864\\
135	-20.5804616645864\\
140	-20.5804616645864\\
145	-20.5804616645864\\
150	-20.5804616645864\\
155	-20.5804616645864\\
160	-20.5804616645864\\
165	-20.5804616645864\\
170	-20.5804616645864\\
175	-20.5804616645864\\
180	-20.5804616645864\\
180	-20.5804616645864\\
};
\addplot [color=mycolor2, dashed, line width=1.0pt, forget plot]
  table[row sep=crcr]{%
-180	-20.5248373530783\\
-175	-20.5248373530783\\
-170	-20.5248373530783\\
-165	-20.5248373530783\\
-160	-20.5248373530783\\
-155	-20.5248373530783\\
-150	-20.5248373530783\\
-145	-20.5248373530783\\
-140	-20.5248373530783\\
-135	-20.5248373530783\\
-130	-20.5248373530783\\
-125	-20.5248373530783\\
-120	-20.5248373530783\\
-115	-20.5248373530783\\
-110	-20.5248373530783\\
-105	-20.5248373530783\\
-100	-20.5248373530783\\
-95	-20.5248373530783\\
-90	-20.5248373530783\\
-85	-20.5248373530783\\
-80	-20.5248373530783\\
-75	-15.3452022466899\\
-70	-10.4295249784559\\
-65	-6.66336907157028\\
-60	-3.63826513089993\\
-55	-1.13854214123021\\
-50	0.962414953278412\\
-45	2.74429137342316\\
-40	4.25967127953546\\
-35	5.54440484570368\\
-30	6.62347204436122\\
-25	7.51450532963494\\
-20	8.23000482532928\\
-15	8.7787768879648\\
-10	9.16688444991136\\
-5	9.39827242116408\\
0	9.47516264692172\\
-0	9.47516264692172\\
5	9.39827242116408\\
10	9.16688444991136\\
15	8.77877688796479\\
20	8.23000482532927\\
25	7.51450532963492\\
30	6.6234720443612\\
35	5.54440484570367\\
40	4.25967127953543\\
45	2.74429137342313\\
50	0.96241495327839\\
55	-1.13854214123023\\
60	-3.63826513089995\\
65	-6.66336907157031\\
70	-10.4295249784559\\
75	-15.3452022466899\\
80	-20.5248373530783\\
85	-20.5248373530783\\
90	-20.5248373530783\\
95	-20.5248373530783\\
100	-20.5248373530783\\
105	-20.5248373530783\\
110	-20.5248373530783\\
115	-20.5248373530783\\
120	-20.5248373530783\\
125	-20.5248373530783\\
130	-20.5248373530783\\
135	-20.5248373530783\\
140	-20.5248373530783\\
145	-20.5248373530783\\
150	-20.5248373530783\\
155	-20.5248373530783\\
160	-20.5248373530783\\
165	-20.5248373530783\\
170	-20.5248373530783\\
175	-20.5248373530783\\
180	-20.5248373530783\\
180	-20.5248373530783\\
};
\addplot [color=mycolor1, line width=1.0pt, forget plot]
  table[row sep=crcr]{%
-180	-22.4346729561032\\
-175	-22.4346729561032\\
-170	-22.4346729561032\\
-165	-22.4346729561032\\
-160	-22.4346729561032\\
-155	-22.4346729561032\\
-150	-22.4346729561032\\
-145	-22.4346729561032\\
-140	-22.4346729561032\\
-135	-22.4346729561032\\
-130	-22.4346729561032\\
-125	-22.4346729561032\\
-120	-22.4346729561032\\
-115	-22.4346729561032\\
-110	-22.4346729561032\\
-105	-22.4346729561032\\
-100	-22.4346729561032\\
-95	-22.4346729561032\\
-90	-22.4346729561032\\
-85	-22.4346729561032\\
-80	-15.4042432406048\\
-75	-8.59821655881423\\
-70	-3.93914608175753\\
-65	-0.50847129965584\\
-60	2.09734428717826\\
-55	4.08710054682537\\
-50	5.57788450370362\\
-45	6.63611282228979\\
-40	7.29564840027879\\
-35	7.56532704389684\\
-30	7.42981840479181\\
-25	6.84364357501799\\
-20	5.7134520369338\\
-15	3.85156426954269\\
-10	0.836709546492705\\
-5	-4.54015866459364\\
0	-13.9031478004824\\
-0	-13.9031478004824\\
5	-4.54015866459366\\
10	0.836709546492703\\
15	3.85156426954269\\
20	5.71345203693381\\
25	6.843643575018\\
30	7.42981840479182\\
35	7.56532704389684\\
40	7.29564840027879\\
45	6.63611282228979\\
50	5.57788450370361\\
55	4.08710054682537\\
60	2.09734428717827\\
65	-0.508471299655836\\
70	-3.93914608175753\\
75	-8.59821655881423\\
80	-15.4042432406048\\
85	-22.4346729561032\\
90	-22.4346729561032\\
95	-22.4346729561032\\
100	-22.4346729561032\\
105	-22.4346729561032\\
110	-22.4346729561032\\
115	-22.4346729561032\\
120	-22.4346729561032\\
125	-22.4346729561032\\
130	-22.4346729561032\\
135	-22.4346729561032\\
140	-22.4346729561032\\
145	-22.4346729561032\\
150	-22.4346729561032\\
155	-22.4346729561032\\
160	-22.4346729561032\\
165	-22.4346729561032\\
170	-22.4346729561032\\
175	-22.4346729561032\\
180	-22.4346729561032\\
180	-22.4346729561032\\
};
\addplot [color=mycolor2, line width=1.0pt, forget plot]
  table[row sep=crcr]{%
-180	-21.2058931810379\\
-175	-21.2058931810379\\
-170	-21.2058931810379\\
-165	-21.2058931810379\\
-160	-21.2058931810379\\
-155	-21.2058931810379\\
-150	-21.2058931810379\\
-145	-21.2058931810379\\
-140	-21.2058931810379\\
-135	-21.2058931810379\\
-130	-21.2058931810379\\
-125	-21.2058931810379\\
-120	-21.2058931810379\\
-115	-21.2058931810379\\
-110	-21.2058931810379\\
-105	-21.2058931810379\\
-100	-21.2058931810379\\
-95	-21.2058931810379\\
-90	-21.2058931810379\\
-85	-21.2058931810379\\
-80	-21.2058931810379\\
-75	-15.1959411849186\\
-70	-10.3277511066267\\
-65	-6.61902462991493\\
-60	-3.6596975135561\\
-55	-1.23204390139299\\
-50	0.792829382090891\\
-45	2.49698110377142\\
-40	3.93536811158544\\
-35	5.14613258273681\\
-30	6.15639932007887\\
-25	6.9857444660588\\
-20	7.64836847259274\\
-15	8.15450301645338\\
-10	8.51133872264149\\
-5	8.72363527794835\\
0	8.79410681896212\\
-0	8.79410681896212\\
5	8.72363527794841\\
10	8.5113387226416\\
15	8.15450301645354\\
20	7.64836847259295\\
25	6.98574446605906\\
30	6.15639932007918\\
35	5.14613258273716\\
40	3.93536811158583\\
45	2.49698110377184\\
50	0.792829382091336\\
55	-1.23204390139251\\
60	-3.65969751355559\\
65	-6.6190246299144\\
70	-10.3277511066262\\
75	-15.1959411849181\\
80	-21.2058931810379\\
85	-21.2058931810379\\
90	-21.2058931810379\\
95	-21.2058931810379\\
100	-21.2058931810379\\
105	-21.2058931810379\\
110	-21.2058931810379\\
115	-21.2058931810379\\
120	-21.2058931810379\\
125	-21.2058931810379\\
130	-21.2058931810379\\
135	-21.2058931810379\\
140	-21.2058931810379\\
145	-21.2058931810379\\
150	-21.2058931810379\\
155	-21.2058931810379\\
160	-21.2058931810379\\
165	-21.2058931810379\\
170	-21.2058931810379\\
175	-21.2058931810379\\
180	-21.2058931810379\\
180	-21.2058931810379\\
};
\end{axis}
\end{tikzpicture}%
    \caption{Directivity of the far-field of the first CM of the modified structure~(\lineleg{mycolor1}), the second CM of the modified structure~(\lineleg{mycolor2}), for $\delta = 0$ and $\sigma = 0.01\lambda_0$ and the first CM of the base structure (\lineleg{mycolor1,dashed}) and the third CM of the base structure (\lineleg{mycolor2,dashed}).}
    \label{patch_modes_modified_vs_base}
\end{figure}

Due to the modification of the structure, the characteristic far-fields of the modified structure are different than the characteristic far-fields of the base structure, as seen in Fig.~\ref{patch_modes_modified_vs_base}. Now, one potential step forward would be to calculate new optimal weighting coefficients in the eigenbasis of the new modified structure to achieve the radiation null. However, as more geometric changes are about to follow, this would mean searching for different weighting coefficients every time the geometry is changed. Therefore, another approach is taken here. Since the modified structure is a substructure of the base structure, the modal transformation matrix is calculated to transform the modal weighting coefficients to the CMs of the base structure. This enables to continue using the calculated desired ratio \mbox{$f_1/f_3$} from the beginning.

\begin{figure}
    \centering
    \pgfkeys{/pgf/number format/.cd,fixed}
    \input{images/patch_mode_ampl_initial}
    \caption{Magnitude of the ratio of the radiated modal weighting coefficients $|f_1/f_3|$ as function of the probe feed position $x_\mathrm{P}$, for a centered inner patch ($\delta = 0$) and a gap of $\sigma = 0.01\lambda_0$.}
    \label{fig:patch_mode_ampl_initial}
\end{figure}

Now, this ratio \mbox{$|f_1/f_3|$} is shown in Fig.~\ref{fig:patch_mode_ampl_initial} for the different port positions $x_\mathrm{P}$. While the desired ratio of the weighting coefficients is \mbox{$|f_1/f_3| = 2.6$}, it is seen that for this configuration of the modified structure, the ratio \mbox{$|f_1/f_3|$} is either very high if the probe-feed is placed on the outer patch or very low if the probe-feed is placed on the inner patch, with a very steep transition in between. The reason for this is that the current of the first mode approximately only flows on the outer patch, while the current of the third mode approximately only flows on the inner patch. This makes it impossible to excite both of them with only a single probe-feed.

\begin{figure}
    \centering
%
%
\definecolor{mycolor1}{rgb}{0.00000,0.44700,0.74100}%
\definecolor{mycolor2}{rgb}{0.85000,0.32500,0.09800}%
\begin{tikzpicture}

\begin{axis}[%
width=2.307in,
height=1.273in,
at={(0.543in,0.501in)},
scale only axis,
xmin=-90,
xmax=90,
xtick={-90, -60, -30,   0,  30,  60,  90},
xlabel style={font=\color{white!15!black}},
xlabel={$\theta\text{ in }{\circ}$},
ymin=-10,
ymax=15,
ylabel style={font=\color{white!15!black}},
ylabel={Directivity in dBi},
axis background/.style={fill=white},
xmajorgrids,
ymajorgrids
]
\addplot [color=mycolor1, dashed, line width=1.0pt, forget plot]
  table[row sep=crcr]{%
-180	-20.5804616645864\\
-175	-20.5804616645864\\
-170	-20.5804616645864\\
-165	-20.5804616645864\\
-160	-20.5804616645864\\
-155	-20.5804616645864\\
-150	-20.5804616645864\\
-145	-20.5804616645864\\
-140	-20.5804616645864\\
-135	-20.5804616645864\\
-130	-20.5804616645864\\
-125	-20.5804616645864\\
-120	-20.5804616645864\\
-115	-20.5804616645864\\
-110	-20.5804616645864\\
-105	-20.5804616645864\\
-100	-20.5804616645864\\
-95	-20.5804616645864\\
-90	-20.5804616645864\\
-85	-20.5804616645864\\
-80	-13.5582012282876\\
-75	-6.75114654338618\\
-70	-2.09095419711052\\
-65	1.34098763323891\\
-60	3.94816296548806\\
-55	5.93927903331796\\
-50	7.43129613998504\\
-45	8.49045958256861\\
-40	9.15038815685999\\
-35	9.41953833541356\\
-30	9.28192083636998\\
-25	8.69075111801327\\
-20	7.54964736528213\\
-15	5.66224454778262\\
-10	2.57277683882174\\
-5	-3.19007117194684\\
0	-20.5804616645864\\
-0	-20.5804616645864\\
5	-3.19007117194684\\
10	2.57277683882174\\
15	5.66224454778262\\
20	7.54964736528214\\
25	8.69075111801326\\
30	9.28192083636998\\
35	9.41953833541356\\
40	9.15038815686\\
45	8.4904595825686\\
50	7.43129613998505\\
55	5.93927903331795\\
60	3.94816296548806\\
65	1.3409876332389\\
70	-2.09095419711052\\
75	-6.75114654338618\\
80	-13.5582012282876\\
85	-20.5804616645864\\
90	-20.5804616645864\\
95	-20.5804616645864\\
100	-20.5804616645864\\
105	-20.5804616645864\\
110	-20.5804616645864\\
115	-20.5804616645864\\
120	-20.5804616645864\\
125	-20.5804616645864\\
130	-20.5804616645864\\
135	-20.5804616645864\\
140	-20.5804616645864\\
145	-20.5804616645864\\
150	-20.5804616645864\\
155	-20.5804616645864\\
160	-20.5804616645864\\
165	-20.5804616645864\\
170	-20.5804616645864\\
175	-20.5804616645864\\
180	-20.5804616645864\\
180	-20.5804616645864\\
};
\addplot [color=mycolor2, dashed, line width=1.0pt, forget plot]
  table[row sep=crcr]{%
-180	-20.5248373530783\\
-175	-20.5248373530783\\
-170	-20.5248373530783\\
-165	-20.5248373530783\\
-160	-20.5248373530783\\
-155	-20.5248373530783\\
-150	-20.5248373530783\\
-145	-20.5248373530783\\
-140	-20.5248373530783\\
-135	-20.5248373530783\\
-130	-20.5248373530783\\
-125	-20.5248373530783\\
-120	-20.5248373530783\\
-115	-20.5248373530783\\
-110	-20.5248373530783\\
-105	-20.5248373530783\\
-100	-20.5248373530783\\
-95	-20.5248373530783\\
-90	-20.5248373530783\\
-85	-20.5248373530783\\
-80	-20.5248373530783\\
-75	-15.3452022466899\\
-70	-10.4295249784559\\
-65	-6.66336907157028\\
-60	-3.63826513089993\\
-55	-1.13854214123021\\
-50	0.962414953278412\\
-45	2.74429137342316\\
-40	4.25967127953546\\
-35	5.54440484570368\\
-30	6.62347204436122\\
-25	7.51450532963494\\
-20	8.23000482532928\\
-15	8.7787768879648\\
-10	9.16688444991136\\
-5	9.39827242116408\\
0	9.47516264692172\\
-0	9.47516264692172\\
5	9.39827242116408\\
10	9.16688444991136\\
15	8.77877688796479\\
20	8.23000482532927\\
25	7.51450532963492\\
30	6.6234720443612\\
35	5.54440484570367\\
40	4.25967127953543\\
45	2.74429137342313\\
50	0.96241495327839\\
55	-1.13854214123023\\
60	-3.63826513089995\\
65	-6.66336907157031\\
70	-10.4295249784559\\
75	-15.3452022466899\\
80	-20.5248373530783\\
85	-20.5248373530783\\
90	-20.5248373530783\\
95	-20.5248373530783\\
100	-20.5248373530783\\
105	-20.5248373530783\\
110	-20.5248373530783\\
115	-20.5248373530783\\
120	-20.5248373530783\\
125	-20.5248373530783\\
130	-20.5248373530783\\
135	-20.5248373530783\\
140	-20.5248373530783\\
145	-20.5248373530783\\
150	-20.5248373530783\\
155	-20.5248373530783\\
160	-20.5248373530783\\
165	-20.5248373530783\\
170	-20.5248373530783\\
175	-20.5248373530783\\
180	-20.5248373530783\\
180	-20.5248373530783\\
};
\addplot [color=mycolor1, line width=1.0pt, forget plot]
  table[row sep=crcr]{%
-180	-22.9406705820077\\
-175	-22.9406705820077\\
-170	-22.9406705820077\\
-165	-22.9406705820077\\
-160	-22.9406705820077\\
-155	-22.9406705820077\\
-150	-22.9406705820077\\
-145	-22.9406705820077\\
-140	-22.9406705820077\\
-135	-22.9406705820077\\
-130	-22.9406705820077\\
-125	-22.9406705820077\\
-120	-22.9406705820077\\
-115	-22.9406705820077\\
-110	-22.9406705820077\\
-105	-22.9406705820077\\
-100	-22.9406705820077\\
-95	-22.9406705820077\\
-90	-22.9406705820077\\
-85	-22.9406705820077\\
-80	-16.6547984915303\\
-75	-9.83566487577377\\
-70	-5.1572774126954\\
-65	-1.69961091886668\\
-60	0.942948830950432\\
-55	2.98226844715776\\
-50	4.54003863779716\\
-45	5.68974886336794\\
-40	6.4764701386967\\
-35	6.92749660485426\\
-30	7.05932941799232\\
-25	6.88443359662693\\
-20	6.42176981288497\\
-15	5.71786279035057\\
-10	4.88734318123556\\
-5	4.16343378258732\\
0	3.8656517039405\\
-0	3.86565170394051\\
5	4.16343378258733\\
10	4.88734318123556\\
15	5.71786279035057\\
20	6.42176981288498\\
25	6.88443359662694\\
30	7.05932941799233\\
35	6.92749660485427\\
40	6.4764701386967\\
45	5.68974886336794\\
50	4.54003863779716\\
55	2.98226844715776\\
60	0.942948830950433\\
65	-1.69961091886668\\
70	-5.1572774126954\\
75	-9.83566487577377\\
80	-16.6547984915303\\
85	-22.9406705820077\\
90	-22.9406705820077\\
95	-22.9406705820077\\
100	-22.9406705820077\\
105	-22.9406705820077\\
110	-22.9406705820077\\
115	-22.9406705820077\\
120	-22.9406705820077\\
125	-22.9406705820077\\
130	-22.9406705820077\\
135	-22.9406705820077\\
140	-22.9406705820077\\
145	-22.9406705820077\\
150	-22.9406705820077\\
155	-22.9406705820077\\
160	-22.9406705820077\\
165	-22.9406705820077\\
170	-22.9406705820077\\
175	-22.9406705820077\\
180	-22.9406705820077\\
180	-22.9406705820077\\
};
\addplot [color=mycolor2, line width=1.0pt, forget plot]
  table[row sep=crcr]{%
-180	-21.6301563726032\\
-175	-21.6301563726032\\
-170	-21.6301563726032\\
-165	-21.6301563726032\\
-160	-21.6301563726032\\
-155	-21.6301563726032\\
-150	-21.6301563726032\\
-145	-21.6301563726032\\
-140	-21.6301563726032\\
-135	-21.6301563726032\\
-130	-21.6301563726032\\
-125	-21.6301563726032\\
-120	-21.6301563726032\\
-115	-21.6301563726032\\
-110	-21.6301563726032\\
-105	-21.6301563726032\\
-100	-21.6301563726032\\
-95	-21.6301563726032\\
-90	-21.6301563726032\\
-85	-21.6301563726032\\
-80	-20.8324173769957\\
-75	-13.9289763390737\\
-70	-9.12626165544896\\
-65	-5.50134578288287\\
-60	-2.64346453640243\\
-55	-0.33379395986275\\
-50	1.55823404195568\\
-45	3.1170970129429\\
-40	4.40106426013069\\
-35	5.45259208893939\\
-30	6.30415056925668\\
-25	6.98159832407512\\
-20	7.50609394047119\\
-15	7.89504676755219\\
-10	8.16240711449274\\
-5	8.31853052619948\\
0	8.36984362739675\\
-0	8.36984362739675\\
5	8.31853052619945\\
10	8.16240711449269\\
15	7.89504676755212\\
20	7.50609394047109\\
25	6.98159832407499\\
30	6.30415056925654\\
35	5.45259208893925\\
40	4.40106426013052\\
45	3.11709701294273\\
50	1.55823404195551\\
55	-0.33379395986293\\
60	-2.64346453640262\\
65	-5.50134578288306\\
70	-9.12626165544915\\
75	-13.9289763390739\\
80	-20.8324173769959\\
85	-21.6301563726032\\
90	-21.6301563726032\\
95	-21.6301563726032\\
100	-21.6301563726032\\
105	-21.6301563726032\\
110	-21.6301563726032\\
115	-21.6301563726032\\
120	-21.6301563726032\\
125	-21.6301563726032\\
130	-21.6301563726032\\
135	-21.6301563726032\\
140	-21.6301563726032\\
145	-21.6301563726032\\
150	-21.6301563726032\\
155	-21.6301563726032\\
160	-21.6301563726032\\
165	-21.6301563726032\\
170	-21.6301563726032\\
175	-21.6301563726032\\
180	-21.6301563726032\\
180	-21.6301563726032\\
};
\end{axis}
\end{tikzpicture}%
    \caption{Directivity of the far-field of the first CM of the modified structure~(\lineleg{mycolor1}), the second CM of the modified structure~(\lineleg{mycolor2}), for $\delta = 0.01\lambda_0$ and $\sigma = 0.01\lambda_0$ and the first CM of the base structure~(\lineleg{mycolor1,dashed}) and the third CM of the base structure~(\lineleg{mycolor2,dashed}).}
    \label{fig:patch_modes_altered2}
\end{figure}

\begin{figure}
    \centering
    \pgfkeys{/pgf/number format/.cd,fixed}
    \input{images/patch_mode_ampl_small_offset}
    \caption{Magnitude of the ratio of the radiated modal weighting coefficients $|f_1/f_3|$ as function of the probe feed position $x_\mathrm{P}$, for an inner patch offset of $\delta = 0.01\lambda_0$ and a gap of $\sigma = 0.01\lambda_0$.}
    \label{fig:patch_mode_ampl_small_offset}
\end{figure}

However, by breaking the symmetry \cite{masek_excitation_2021,peitzmeier_systematic_2022} and shifting the inner patch by an offset $\delta$ in the $x$\nobreakdash-direction, as shown in Fig.~\ref{fig:patch_scattering_variation2}, the modes on the two substructures can be coupled. The coupling can be seen for example in the directivity of the eigenmodes of the structure, as shown in Fig.~\ref{fig:patch_modes_altered2}.

A consequence of the coupling is that both modes can now be excited by a single-probe feed. The ratio of the magnitude of the excited weighting coefficients for this new configuration is again shown in Fig.~\ref{fig:patch_mode_ampl_small_offset}. It is seen that the first mode is excited more than the third mode if the probe-feed is placed on the outer patch, and vice versa if the probe feed is placed on the inner patch. Apart from this, the ratio \mbox{$|f_1/f_3|$} is approximately constant, no matter where the probe-feed is placed on the patch. Therefore, in the following, a fixed probe-feed position at \mbox{$x_\mathrm{P} = -0.19\lambda_0$} is considered.



\begin{figure}
    \centering
    \pgfplotsset{scaled ticks=false}
    \pgfkeys{/pgf/number format/.cd,fixed}
    \input{images/patch_mode_ampl_phase_vs_offset}
    \caption{Magnitude and phase of the ratio of the radiated modal weighting coefficients \mbox{$f_1/f_3$} as function of the offset of the inner patch $\delta$ for a gap \mbox{$\sigma = 0.01\lambda_0$} and the probe feed position \mbox{$x_\mathrm{P} = -0.19\lambda_0$}.}
    \label{fig:patch_mode_ampl_phase_vs_offset}
\end{figure}

\begin{figure}
    \centering
    \pgfplotsset{scaled ticks=false}
    \pgfkeys{/pgf/number format/.cd,fixed,precision=3}
    \input{images/patch_mode_ampl_phase_vs_gap}
    \caption{Magnitude and phase of the ratio of the radiated modal weighting coefficients \mbox{$\angle f_1/f_3$} as function of the gap $\sigma$ for the selected offset \mbox{$\delta = 0.02\lambda_0$} and the probe feed position \mbox{$x_\mathrm{P} = -0.19\lambda_0$}.}
    \label{fig:patch_mode_ampl_phase_vs_gap}
\end{figure}

Now, in order to control the coupling between the two modes, the offset $\delta$ is varied, as shown in Fig.~\ref{fig:patch_mode_ampl_phase_vs_offset}. Thereby, it is especially seen that only the magnitude of the ratio is changed, while the phase approximately remains constant. In order to achieve the desired \mbox{$|f_1/f_3| = 2.6$}, the offset is chosen to be \mbox{$\delta = 0.02\lambda_0$} now.

In order to control the phase on the other hand, the gap is varied, as shown in Fig~\ref{fig:patch_mode_ampl_phase_vs_gap}. Thereby, it is especially seen that only the phase of the ratio is changed, while the magnitude remains approximately constant. In order to achieve the desired \mbox{$\angle f_1/f_3 \approx 90^\circ$}, the gap size of \mbox{$\sigma = 0.015\lambda_0$} is chosen.


Now, since the modal configuration is not significantly altered by the port position, as seen earlier, it is tuned to achieve an input impedance matching of better than \mbox{$|\mathbf{s}_{11}| < -\SI{10}{\deci\bel}$} for a reference impedance of $\SI{50}{\ohm}$, which yields a port position of \mbox{$x_\mathrm{P} = -0.13\lambda_0$}. The realized far-field for this configuration is shown in Fig.~\ref{fig:patch_realized2}. It is seen that the goal to have a radiation null at \mbox{$\theta = -15^\circ$} is indeed achieved by this configuration and the radiation pattern looks very similar to the synthetic pattern that was shown in Fig.~\ref{fig:patch_modes2}.

Finally, it is noted that the characteristic far-field patterns of the modified structure in its eigenbasis vary significantly with respect to the parameter changes as it was seen in Fig.~\ref{fig:patch_modes_altered2}. In order to eliminate the complexity of a varying far-field, the modal weighting coefficients were transformed from the eigenbasis of the modified structure into modal weighing coefficients of the CMs of the base structure, as proposed earlier in this paper. Without doing this, the methodology that was shown in this example would have been impossible, since the interpretation of the modal weighting coefficients would have differed for every parameter step, making it difficult to interpret the results.

Summarized, the design of a patch antenna with a radiation null in a specific direction was investigated using a fixed CM basis. Due to the fixed basis, the investigation could be reduced to the analysis of just the modal weighting coefficients, in favor of a decomposition where fields and weighting coefficients would both change. Furthermore, the example also showed that the proposed theory is compatible with the use of an an infinitely extended grounded dielectric slab as background structure.

\begin{figure}
    \centering
%
%
\definecolor{mycolor1}{rgb}{0.63500,0.07800,0.18400}%
\definecolor{mycolor2}{rgb}{0.92900,0.69400,0.12500}%
\begin{tikzpicture}

\begin{axis}[%
width=2.307in,
height=1.273in,
at={(0.543in,0.501in)},
scale only axis,
xmin=-90,
xmax=90,
xtick={-90, -60, -30,   0,  30,  60,  90},
xlabel style={font=\color{white!15!black}},
xlabel={$\theta\text{ in }{\circ}$},
ymin=-10,
ymax=15,
ylabel style={font=\color{white!15!black}},
ylabel={Directivity in dBi},
axis background/.style={fill=white},
axis x line*=bottom,
axis y line*=left,
xmajorgrids,
ymajorgrids
]
\addplot [color=mycolor1, line width=1.0pt, forget plot]
  table[row sep=crcr]{%
-180	-19.8818298875784\\
-175	-19.8818298875784\\
-170	-19.8818298875784\\
-165	-19.8818298875784\\
-160	-19.8818298875784\\
-155	-19.8818298875784\\
-150	-19.8818298875784\\
-145	-19.8818298875784\\
-140	-19.8818298875784\\
-135	-19.8818298875784\\
-130	-19.8818298875784\\
-125	-19.8818298875784\\
-120	-19.8818298875784\\
-115	-19.8818298875784\\
-110	-19.8818298875784\\
-105	-19.8818298875784\\
-100	-19.8818298875784\\
-95	-19.8818298875784\\
-90	-19.8818298875784\\
-85	-19.8818298875784\\
-80	-14.3277176498086\\
-75	-8.89436374407686\\
-70	-5.31615343318481\\
-65	-2.70488889743156\\
-60	-0.74641502831309\\
-55	0.687891579124176\\
-50	1.65062497683677\\
-45	2.15048253222517\\
-40	2.15699050387621\\
-35	1.58709958147059\\
-30	0.261873940695885\\
-25	-2.22121190176076\\
-20	-6.89975585407385\\
-15	-12.5568667525336\\
-10	-4.96050359485057\\
-5	0.409569071450185\\
0	3.80007618752028\\
-0	3.80007618752028\\
5	6.13534261612439\\
10	7.7876346390425\\
15	8.93500431598526\\
20	9.67336231389349\\
25	10.0574580491679\\
30	10.1181701124216\\
35	9.87033429172362\\
40	9.31606481649258\\
45	8.44539128816613\\
50	7.23468705680413\\
55	5.64246034304959\\
60	3.60090188192216\\
65	0.999030191741887\\
70	-2.35423869649024\\
75	-6.82753608677256\\
80	-13.3079319430656\\
85	-19.8818298875784\\
90	-19.8818298875784\\
95	-19.8818298875784\\
100	-19.8818298875784\\
105	-19.8818298875784\\
110	-19.8818298875784\\
115	-19.8818298875784\\
120	-19.8818298875784\\
125	-19.8818298875784\\
130	-19.8818298875784\\
135	-19.8818298875784\\
140	-19.8818298875784\\
145	-19.8818298875784\\
150	-19.8818298875784\\
155	-19.8818298875784\\
160	-19.8818298875784\\
165	-19.8818298875784\\
170	-19.8818298875784\\
175	-19.8818298875784\\
180	-19.8818298875784\\
180	-19.8818298875784\\
};
\addplot [color=mycolor2, dashed, line width=1.0pt, forget plot]
  table[row sep=crcr]{%
-180	-19.9949273065973\\
-175	-19.9949273065973\\
-170	-19.9949273065973\\
-165	-19.9949273065973\\
-160	-19.9949273065973\\
-155	-19.9949273065973\\
-150	-19.9949273065973\\
-145	-19.9949273065973\\
-140	-19.9949273065973\\
-135	-19.9949273065973\\
-130	-19.9949273065973\\
-125	-19.9949273065973\\
-120	-19.9949273065973\\
-115	-19.9949273065973\\
-110	-19.9949273065973\\
-105	-19.9949273065973\\
-100	-19.9949273065973\\
-95	-19.9949273065973\\
-90	-19.9949273065973\\
-85	-19.9949273065973\\
-80	-19.1124667072887\\
-75	-12.3638163760139\\
-70	-7.78832935333371\\
-65	-4.47250966106746\\
-60	-2.01997571004443\\
-55	-0.232534454513093\\
-50	0.990775628021898\\
-45	1.69094860766774\\
-40	1.85960498434206\\
-35	1.42999147566662\\
-30	0.23628263741233\\
-25	-2.12016077647387\\
-20	-6.83693711094402\\
-15	-17.9724594075047\\
-10	-6.71813373210114\\
-5	-0.387862482448728\\
0	3.30380176813402\\
-0	3.30380176813402\\
5	5.78649573610075\\
10	7.52727171374008\\
15	8.7334993511361\\
20	9.51286019730485\\
25	9.92554799522163\\
30	10.0050726934027\\
35	9.76738792971454\\
40	9.21470711541034\\
45	8.33625233731746\\
50	7.10651216218584\\
55	5.48050469990689\\
60	3.38415619733337\\
65	0.69498485733576\\
70	-2.79966177158691\\
75	-7.5060855622572\\
80	-14.3448677952917\\
85	-19.9949273065973\\
90	-19.9949273065973\\
95	-19.9949273065973\\
100	-19.9949273065973\\
105	-19.9949273065973\\
110	-19.9949273065973\\
115	-19.9949273065973\\
120	-19.9949273065973\\
125	-19.9949273065973\\
130	-19.9949273065973\\
135	-19.9949273065973\\
140	-19.9949273065973\\
145	-19.9949273065973\\
150	-19.9949273065973\\
155	-19.9949273065973\\
160	-19.9949273065973\\
165	-19.9949273065973\\
170	-19.9949273065973\\
175	-19.9949273065973\\
180	-19.9949273065973\\
180	-19.9949273065973\\
};
\end{axis}
\end{tikzpicture}%
    \caption{Directivity of the realized far-field for the modified structure for \mbox{$w_2 = 0.235 \lambda_0$}, \mbox{$\delta = 0.02\lambda_0$} and \mbox{$\sigma = 0.015\lambda_0$} with a probe-feed placed at \mbox{$x_\mathrm{P} = -0.13\lambda_0$}, simulated with a full-wave simulation with actual probe-feed (\lineleg{mycolor1}) and with the used assumption of an impressed current filament model according to \cite{yang_computing_2016} (\lineleg{mycolor2,dashed}).}
    \label{fig:patch_realized2}
\end{figure}

\section{Discussion}
\label{sec:discussion}

Traditionally, CMs have only been used to analyze the scattering of the structure whose scattering operator they diagonalize. However, the formalism proposed in this paper shows that CMs are not limited to this case. CMs can also be used to describe scattering of other objects whose surfaces are a subset of the structure that was used to define the CMs, at least in the case of non-closed conducting structures which was investigated in this paper. This enables the analysis of geometrical changes of scattering objects in a common CM basis, as it was seen in the examples in this paper.

Before this contribution, spherical wave functions \cite{hansen_spherical_1988} were often used to describe scattering of different objects in a common basis \cite{wasylkiwskyj_scattering_1970,rubio_generalized-scattering-matrix_2005,izquierdo_spherical-waves-based_2012,berkelmann_antenna_2022,jesus_rubio_analysis_2022}. However, spherical wave functions are by their definition only suitable to be used in free-space, which means that they can not be used to describe scenarios with other background media such as the grounded dielectric slab in the third example. Furthermore, they contain a lot of basis functions that do not interact with planar structures, making them an inefficient representation in terms of the number of coefficients to store. Finally, it should be noted that the use of spherical wave functions actually constitutes a special case of the proposed theory, since spherical wave functions are the CMs of a sphere in free-space. 

Furthermore, it should be noted that in this paper we have focused on the strict case that the proposed method only converges if the surface of the scattering object is a subset of the surface used to define the CMs. However, during the development of the paper, we have also found that the model can provide a good approximation even if this condition is violated only to a certain degree. Still, since this is outside of the scope of this paper, we have not investigated in which cases and to what extend this approximation is accurate.

Finally, it is noted that the proposed theory should not be confused with the so-called substructure characteristic modes, as e.g. discussed in \cite{ethier_sub-structure_2012,ma_influence_2019,gustafsson_theory_2024,shi_computation_2024,shi_scattering-based_2024}. While substructure characteristic modes also deal with two structures of which one structure is a substructure of the other, they define an eigenvalue problem based on the scattering behavior of both structures. Here, instead, the characteristic fields of the superstructure are used to describe the scattering behavior of just the substructure, without taking the scattering behavior of the superstructure into account.

\section{Conclusion}
\label{sec:conclusion}

A formalism that enables to use the CMs of one structure to analyze the scattering of another structure is proposed. The characteristic fields of the first structure are thereby used as incident field to the second structure, which requires the use of a new method of moments type matrix that we call cross radiation matrix. With the help of this matrix, the scattering of the second structure is represented using a generally non-diagonal perturbation or scattering matrix in the CMs of the first structure. In a similar manner, a mapping between the CMs of both structures is established in the form of a transformation matrix.

\annotateChangesStart Three \annotateChangesEnd examples demonstrate how this formalism can be used to describe different geometries in a common mathematical basis. The results are computationally accurate and illustrate that the formalism can be used in various scenarios, e.g. in the far-field, in the near-field, with a grounded dielectric slab. This makes the formalism a good candidate for the use in computational methods investigating scattering with varying object geometries. Furthermore, in the \annotateChangesStart third \annotateChangesEnd example, it is seen that the CMs of a base structure can sometimes also be a more intuitive basis than the CMs of the structure itself for the design task that should be accomplished.

\appendices

\section{The Cross Radiation Matrix} 
\label{sec:interpretation_RAB}

The matrix $\mathbf{R}^{\mathrm{AB}}$ was derived mathematically by testing the real part of the scattered field of the basis functions at the test functions of another structure that is allowed to have a different discretization. However, we propose to call it cross radiation matrix, since the formula of the matrix is very similar to the real part of the impedance matrix of the method of moments \annotateChangesStart
\begin{equation} 
\label{eq:re_z_eq_r}
\operatorname{Re}\{ \mathbf{Z}\} = \mathbf{R},
\end{equation}
\annotateChangesEnd which is also often called radiation matrix \cite{gustafsson_unified_2022-1}:
\begin{equation}
\label{eq:R0}
    \mathbf{R} = \left[\int_\Omega\int_\Omega\widetilde{\mathbf{\psi}}_{\mu}(\mathbf{r}) \operatorname{Re}\left\{\mathbf{G}(\mathbf{r}, \mathbf{r}^\prime) \right\} \,\mathbf{\psi}_{m}(\mathbf{r}^\prime)\, \mathrm{d}S \,\mathrm{d}S^\prime \right]_{\mu m}.
\end{equation}

The only difference between \eqref{eq:RBA} and \eqref{eq:R0} is that the basis functions and the test functions are allowed to be from a different structure for the cross radiation matrix. In fact, if the structures A and B are equal and equally discretized, then
\begin{equation}
    \mathbf{R} = \mathbf{R}^{\mathrm{AB}} =  \mathbf{R}^{\mathrm{BA}}.
\end{equation}

Or, in another case, if the structure B is a substructure of structure A, and their discretization is equal, then the cross-radiation matrix $\mathbf{R}^{\mathrm{BA}}$ is a submatrix of the radiation matrix of structure A $\mathbf{R}^{\mathrm{A}}$.

\section{Coherence with Existing CM Definition}
\label{sec:coherence_with_existing_cm_def}

It may be questioned if the proposed formalism is still coherent with the original formulation of a scattering problem in CMs as described in \cite{harrington_theory_1971}. In the following, it is shown that this is the case. Therefore, it is assumed that structures A and B both describe the same object with the same discretization and it is proven that the perturbation that is calculated according to the proposed formalism matches with the one mentioned in \cite{harrington_theory_1971}.

\annotateChangesStart The \annotateChangesEnd inverse of the impedance matrix $\mathbf{Z}$ \annotateChangesStart is given by \annotateChangesEnd
\begin{equation}
\label{eq:inverse_z}
    \mathbf{Z}^{-1} = \mathbf{I}_{\mathrm{CM}} \left(\mathbf{I}+ \mathrm{j} \mathbf{\Lambda}\right)^{-1} \mathbf{I}_{\mathrm{CM}}^{\mathrm{T}}.
\end{equation}
By using \eqref{eq:inverse_z}, the perturbation matrix $\mathbf{P} = \mathbf{P}^{\mathrm{ABA}} = \mathbf{P}^{\mathrm{BBB}}$ according to \eqref{eq:p_aba_mom_definition}
\begin{equation}
\mathbf{P} = - \,\mathbf{U}^{\mathrm{T}} \mathbf{Z} ^{-1} \mathbf{U}
\end{equation}
\annotateChangesStart can be formulated as:
\begin{equation}
    \mathbf{P} = - \,\mathbf{U}^{\mathrm{T}} \, \mathbf{I}_{\mathrm{CM}} \left(\mathbf{I}+ \mathrm{j} \mathbf{\Lambda}\right)^{-1} \, \mathbf{I}_{\mathrm{CM}}^{\mathrm{T}} \mathbf{U}.
\end{equation}
Furthermore, by using \eqref{eq:q_ab_definition_mom},
\begin{equation}
    \mathbf{P} = - \,\mathbf{I}_{\mathrm{CM}}^\mathrm{T} \mathbf{R}\, \mathbf{I}_{\mathrm{CM}} \left(\mathbf{I}+ \mathrm{j} \mathbf{\Lambda}\right)^{-1} \mathbf{I}_{\mathrm{CM}}^{\mathrm{T}} \mathbf{R} \, \mathbf{I}_{\mathrm{CM}},
\end{equation}
is obtained, which can be simplified using 
\eqref{eq:cm_normalization} and \eqref{eq:re_z_eq_r} as:
\begin{equation}
    \mathbf{P} = - \left(\mathbf{I}+ \mathrm{j} \mathbf{\Lambda}\right)^{-1} = - \operatorname{diag} \frac{1}{1 + \mathrm{j}\lambda_n}.
\end{equation}
\annotateChangesEnd This matrix is the diagonal matrix \eqref{eq:p_diag} that was also derived in \cite{harrington_theory_1971} and therefore shows that the proposed framework is in-line with the existing theory.








\ifCLASSOPTIONcaptionsoff
  \newpage
\fi



\bibliographystyle{IEEEtran}
\bibliography{IEEEabrv,lit}

\begin{IEEEbiography}[{\includegraphics[width=1in,height=1.25in,clip,keepaspectratio]{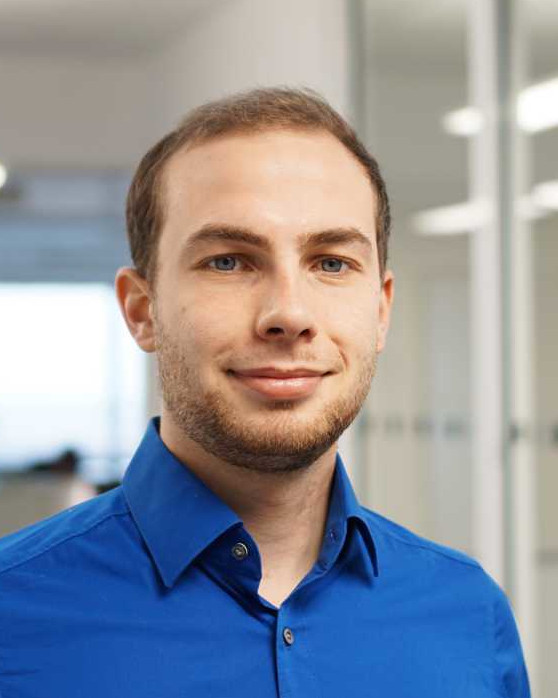}}]{Leonardo Mörlein}
Leonardo Mörlein (Graduate Student Member, IEEE) was born in 1994 in Würzburg, Germany. He received the B.Sc. and M.Sc. degrees in electrical engineering from Leibniz University Hannover, Hannover, Germany, in 2017 and 2020, respectively. He is currently a Research Assistant with the Institute of Microwave and Wireless Systems, Leibniz University Hannover. His current research focuses on the use of multi-port multi-mode antennas in beamforming scenarios. Further research interests include antenna integration, the use of modal decompositions and channel modeling.
\end{IEEEbiography}

\begin{IEEEbiography}[{\includegraphics[width=1in,height=1.25in,clip,keepaspectratio]{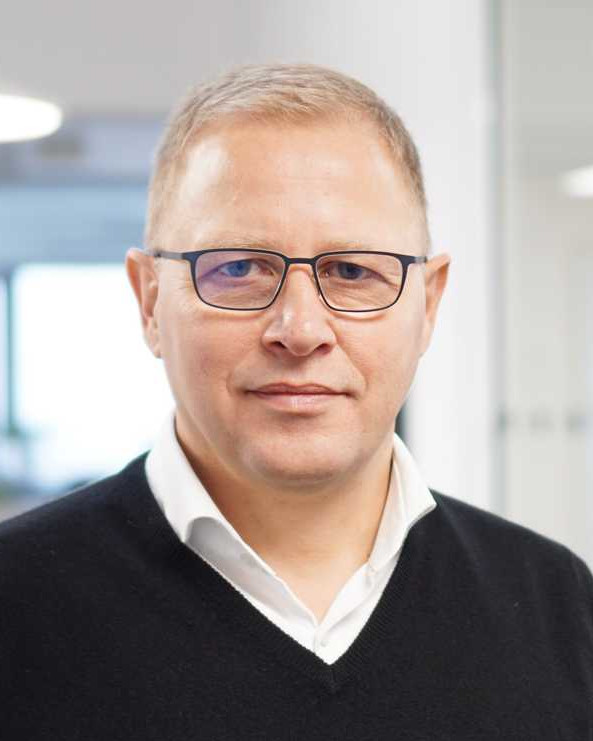}}]{Dirk Manteuffel} (Member, IEEE) was born in Issum, Germany, in 1970. He received the Dipl.-Ing. and Dr.-Ing. degrees in electrical engineering from the University of Duisburg–Essen, Duisburg, Germany, in 1998 and 2002, respectively.

From 1998 to 2009, he was with IMST, Kamp-Lintfort, Germany. As a Project Manager, he was responsible for industrial antenna development and advanced projects in the field of antennas and electromagnetic (EM) modeling. From 2009 to 2016, he was a Full Professor of wireless communications at Christian-Albrechts-University, Kiel, Germany. Since June 2016, he has been a Full Professor and the Executive Director of the Institute of Microwave and Wireless Systems, Leibniz University Hannover, Hannover, Germany. His research interests include electromagnetics, antenna integration and EM modeling for mobile communications and biomedical applications.

Dr. Manteuffel was a director of the European Association on Antennas and Propagation from 2012 to 2015. He served on the Administrative Committee (AdCom) of IEEE Antennas and Propagation Society from 2013 to 2015 and as an Associate Editor of the IEEE Transactions on Antennas and Propagation from 2014 to 2022. Since 2009 he has been an appointed member of the committee "Antennas" of the German VDI-ITG.
\end{IEEEbiography}





\end{document}